\documentclass[11pt]{article} 
\usepackage{graphicx,amssymb,mathrsfs,array,multirow}  
\begin{document}  
 
\vskip 30pt

\begin{center}  
{\Large{\bf A4-based see-saw model for realistic neutrino masses
and mixing}}\\
\vspace*{1cm}  
\renewcommand{\thefootnote}{\fnsymbol{footnote}}  
{ {\sf Soumita Pramanick${}$\footnote{email: soumitapramanick5@gmail.com}},
{\sf Amitava Raychaudhuri${}$\footnote{email: palitprof@gmail.com}} 
} \\  
\vspace{10pt}  
{\small  {\em Department of Physics, University of Calcutta,  
92 Acharya Prafulla Chandra Road, Kolkata 700009, India}}
\normalsize

\end{center}  

\begin{abstract} 

We present an $A4$-based model where neutrino masses arise from a
combination of see-saw mechanisms. The model is motivated by
several small mixing and mass parameters indicated by the data.
These are $\theta_{13}$, the solar mass splitting, and the small
deviation of  $\theta_{23}$ from maximal mixing (= $\pi/4$). We
take  the above as indications that at some level the small
quantities are well-approximated by zero. In particular the
mixing angles, to a zero order, should be either 0 or $\pi/4$.
Accordingly, in this model the Type-II see-saw dominates and
generates the larger atmospheric mass splitting and sets
$\theta_{23} = \pi/4$.   The other mixing angles are vanishing as
is  the solar splitting. We show how the $A4$ assignment for the
lepton doublets leads to this form.  We also specify the $A4$
properties of the right-handed neutrinos which result in a
smaller Type-I see-saw contribution that acts  as a perturbation
and shifts the angles $\theta_{12}$ and $\theta_{13}$ 
into the correct range and  the desired value of $\Delta
m^2_{solar}$ is produced.  The $A4$ symmetry results in
relationships between these quantities as well as with a small
deviation of $\theta_{23}$ from $\pi/4$.  If the right-handed
neutrino mass matrix, $M_R$, is chosen real then there is no
leptonic CP-violation and only Normal Ordering is admissible. If
$M_R$ is  complex then Inverted Ordering is also allowed with the
proviso that the CP-phase, $\delta$, is large, i.e., $\sim
\pi/2$ or $-\pi/2$.  The preliminary results from NO$\nu$A
favouring Normal Ordering and $\delta$ near $-\pi/2$ imply
quasi-degenerate neutrino masses in this model.

\vskip 5pt \noindent  
\texttt{PACS No:~ 14.60.Pq}  \\  
\texttt{Key Words:~~Neutrino mixing, $\theta_{13}$, Solar
splitting, A4, see-saw}
\end{abstract}  

\renewcommand{\thesection}{\Roman{section}} 
\setcounter{footnote}{0} 
\renewcommand{\thefootnote}{\arabic{footnote}} 
\noindent

\section{Introduction}
Many neutrino oscillation experiments have established that
neutrinos are massive and non-degenerate and that the flavour
eigenstates are not identical with the mass eigenstates. For the
three neutrino paradigm the two independent   mass square
splittings are the solar ($\Delta m^2_{solar}$) and the
atmospheric ($\Delta m^2_{atmos}$). The mass and flavour bases
are related through the Pontecorvo, Maki, Nakagawa, Sakata --
PMNS -- matrix usually parametrized as:
\begin{eqnarray}
U = \left(
          \begin{array}{ccc}
          c_{12}c_{13} & s_{12}c_{13} & s_{13}e^{-i\delta}  \\
 -c_{23}s_{12} + s_{23}s_{13}c_{12}e^{i\delta} & c_{23}c_{12} +
s_{23}s_{13}s_{12}e^{i\delta}&  -s_{23}c_{13}\\
 - s_{23}s_{12} - c_{23}s_{13}c_{12}e^{i\delta}&  s_{23}c_{12} -
c_{23}s_{13}s_{12}e^{i\delta} & c_{23}c_{13} \end{array} \right)
\;\;,
\label{PMNS}
\end{eqnarray}
where $c_{ij} = \cos \theta_{ij}$ and $s_{ij} = \sin \theta_{ij}$. 

The recent measurement of a non-zero value for $\theta_{13}$ \cite{t13},
which is small compared to the other mixing angles, has led to a
flurry of activity in developing neutrino mass models which
incorporate this feature. Earlier we had demonstrated \cite{br}
that a direction which bears exploration is whether two small
quantities in the neutrino sector, namely, $\theta_{13}$ and the
ratio, $R \equiv \Delta m^2_{solar}/\Delta m^2_{atmos}$,  could
in fact be related to each other, both resulting from a small
perturbation.  Subsequently we had shown \cite{pr} that it is
possible to envisage scenarios where only the larger $\Delta
m^2_{atmos}$ and $\theta_{23} = \pi/4$ are present in a basic
structure of neutrino mass and mixing and the rest of the
quantities, namely,  $\theta_{13}, \theta_{12}$, the deviation
of $\theta_{23}$ from $\pi/4$, and $\Delta m^2_{solar}$ all have
their origin in a smaller see-saw induced
perturbation\footnote{Early work on neutrino mass models where
some variables are much smaller than others can be found in
\cite{old}.}.  Obviously, this gets reflected in constraints on
the measured quantities. A vanishing $\theta_{13}$ follows
rather easily from certain symmetries and indeed many of the
newer models are based on perturbations of such structures
\cite{models, LuhnKing}.

Encouraged by the success of this program we present here a model
based on the group $A4$ which relies on the see-saw mechanism
\cite{seesaw} in which  the lightest neutrino mass, $m_0$, the see-saw
scale and one other parameter  determine $\theta_{13}$, $R$,
$\theta_{12}$, and the deviation of $\theta_{23}$ from $\pi/4$.
If this last parameter is complex then the CP-phase $\delta$ is
also a prediction. Here, the atmospheric mass splitting is taken as
an input which together with the lightest neutrino mass
completely defines the unperturbed mass matrix generated by the
Type-II see-saw.  The size of the perturbation is determined
by the Type-I see-saw and is of the form $m_D^2/m_R$ where
$m_D$ and $m_R$ respectively are the scale of the Dirac and
right-handed Majorana mass terms.

After  a brief summary of the $A4$ group properties and the
structure of the model in the following section we describe the
implications of the model in the next section. The comparison of
this model with the experimental data appears next. We end with
our conclusions.  The model has a rich scalar field content.  In
an Appendix we discuss the $A4$ invariant scalar potential and
the conditions under which the desired potential minimum can be
realized.

It should be noted that neutrino mass models based on $A4$ have
also been investigated earlier \cite{A4mr, A4af, otherA4}. In a majority 
of them the neutrino mass matrix is obtained from a Type-II
see-saw and the earlier emphasis was on obtaining tribimaximal
mixing. Recent work has focussed on obtaining more realistic
mixing patterns \cite{newMa} sometimes taking recourse to
breaking of $A4$ symmetry \cite{tanimoto}. Our work is unique in
two respects. Firstly, it uses a combination of Type-II and Type-I
see-saw mechanisms where the former yields mixing angles which are
either vanishing ($\theta_{12}$ and $\theta_{13}$) or maximal --
i.e., $\pi/4$  -- ($\theta_{23}$) while keeping the solar
splitting absent. This kind of a scenario has not been considered
before. The Type-I see-saw acting as a perturbation
results in a non-zero CP-phase and realistic mixing angles while
at the same time creating the correct solar splitting. Secondly, all this
is accomplished keeping the $A4$ symmetry intact. 

\section{The Model}

$A4$ is the group of even permutations of four objects comprising
of 12 elements which can be generated using the two basic
permutations $S$ and $T$  satisfying $S^2=T^3=(ST)^3=\mathbb{I}$.
The group has four inequivalent irreducible representations one
of 3 dimension and three of 1 dimension denoted by $1, 1'$ and
$1''$. The one-dimensional representations are all singlets under
$S$ and transform as 1, $\omega$, and $\omega^2$ under $T$
respectively, where $\omega$ is the cube root of unity.  Thus $1'
\times 1'' = 1$. 

For the three-dimensional representation
\begin{equation}
S=\pmatrix{1 & 0 & 0 \cr 0 & -1 & 0 \cr 0 & 0 & -1}\ \ \ \ {\rm and} \ \ \ \
T=
\pmatrix{0 & 1 & 0 \cr
0 & 0 & 1 \cr
1 & 0 & 0} \;\;.
\label{ST3}
\end{equation}
This representation satisfies the product rule
\begin{equation}
3\otimes3=1 \oplus 1' \oplus 1'' \oplus 3 \oplus 3 \;\;.
\label{A43x3}
\end{equation}
The two triplets arising from the product
of $3_a \equiv {a_i}$ and $3_b \equiv {b_i}$, where 
$i=1,2,3$, can be identified as $3_c \equiv {c_i}$
and $3_d \equiv {d_i}$ with 
\begin{eqnarray}
c_i &=& \left(\frac{a_2 b_3 + a_3 b_2}{2},\frac{a_3 b_1 + a_1 b_3}{2},
\frac{a_1 b_2 + a_2 b_1}{2}\right)   \;\;, \;\;{\rm or,}
\;\; c_i \equiv \alpha_{ijk} a_j b_k \;\;, \nonumber \\
d_i &=& \left(\frac{a_2 b_3 - a_3 b_2}{2},\frac{a_3 b_1 - a_1 b_3}{2},
\frac{a_1 b_2 - a_2 b_1}{2}\right)   \;\;, \;\;{\rm or,}
\;\; d_i \equiv \beta_{ijk} a_j b_k \;\;,\;\; (i,j,k, {\rm are ~cyclic})\;\;.
\label{3x3to3}
\end{eqnarray}
In this notation the one-dimensional
representations in the $3\otimes3$ product can be written as: 
\begin{eqnarray}
1 &=& a_1b_1+a_2b_2+a_3b_3 \equiv \rho_{1ij}a_ib_j \;\;,\nonumber \\  
1'&=& a_1b_1+\omega^2a_2b_2+\omega a_3b_3 \equiv \rho_{3ij}a_ib_j
\;\;,  \nonumber \\
1'' &=& a_1b_1+\omega a_2b_2+\omega^2a_3b_3 \equiv \rho_{2ij}a_ib_j \;\;. 
\label{3x3to1}
\end{eqnarray}
Further details of the group
$A4$ are available in the literature \cite{A4mr, A4af}.

\begin{table}[t]
\begin{center}
\begin{tabular}{|c|c|c|c|c|}
\hline
Fields & Notations & $A4$ & $SU(2)_L$ ($Y$) & $L$   \\ 
 \hline
 & & &  &   \\
Left-handed leptons&$(\nu_i,l_i)_L$&3&2 (-1)& 1 \\
 & & &   &  \\
\hline
 & & &   &  \\
 &$l_{1R} $ &1& &   \\
 & & & &    \\
\cline{2-3}
 & & & &    \\
Right-handed charged leptons & $l_{2R} $&$1'$&1 (-2) & 1
\\
 & & &  &   \\
\cline{2-3}
 & & &  &   \\
 & $l_{3R} $&$1''$& &     \\
 & & &  &   \\
\hline
Right-handed  neutrinos& $N_{iR} $&3&1 (0) & -1 \\
\hline
\end{tabular}
\end{center}
\caption{\em The fermion content of the model. The transformation
properties under $A4$ and $SU(2)_L$ are shown. The hypercharge of the fields,
$Y$, and their lepton number, $L$, are also indicated.}
\label{tab1f}
\end{table}

In the proposed model the left-handed lepton doublets of the
three flavours are assumed to form an $A4$ triplet while the
right-handed charged leptons are taken as $1 (e_R)$, $1'
(\mu_R)$, and $1'' (\tau_R)$ under $A4$.  The remaining leptons,
the right-handed neutrinos, form an $A4$ triplet\footnote{We
closely follow the notation of \cite{A4mr}.}.  The lepton content
of the model with the $A4$ and $SU(2)_L$ properties as well as
the lepton number assignments is shown in Table \ref{tab1f}. 
Note that the right-handed neutrinos are assigned lepton number
-1. This choice is made to ensure that the neutrino Dirac mass
matrix takes a form proportional to the identity matrix, as we
remark in the following. The assignment of $A4$ quantum
numbers of the leptons is by no means unique. The entire list of
options for this have been catalogued  in \cite{rb}. Our choice
corresponds to class B of \cite{rb}. We do not discuss the
extension of this model to the quark sector\footnote{For
$A4$-based models dealing with the quark sector see, for example,
\cite{quarks1} and \cite{quarks2}.}.

\begin{table}[t]
\begin{center}
\begin{tabular}{|c|c|c|c|c|c|}
\hline
Purpose & Notations & $A4$ & $SU(2)_L$ & $L$ & $vev$ \\ 
 &  &  & ($Y$) & &  \\  \hline
 & & & & &  \\
Charged fermion mass& $\Phi=\pmatrix{\phi_1^+ & \phi_1^0\cr
\phi_2^+& \phi_2^0\cr\phi_3^+& \phi_3^0}$&3&2 (1) &0&$\langle\Phi\rangle=
\frac{v}{\sqrt{3}} \pmatrix{0 & 1 \cr 0 & 1 \cr 0 & 1 }$ \\
 & & & &  &  \\
 \hline
 & & & & &  \\
 Neutrino Dirac mass& $\eta=(\eta^0 ,\eta^- )$
&1&2 (-1)&2& $\langle\eta\rangle=\pmatrix{0,u}$
\\
 & & & & &  \\
\hline
 & & & &  & \\
Type-I see-saw mass& $\hat{\Delta}^L=\pmatrix{
\hat{\Delta}^{++}_1 & \hat{\Delta}^{+}_1 & \hat{\Delta}^0_1 \cr
 \hat{\Delta}^{++}_2 & \hat{\Delta}^{+}_2 & \hat{\Delta}^0_2 \cr
 \hat{\Delta}^{++}_3 & \hat{\Delta}^{+}_3 & \hat{\Delta}^0_3 }^L$
&3&3 (2)&-2&$\langle \hat{\Delta}^L \rangle= v_L \pmatrix{0
& 0 & 1  \cr
0 & 0 & 0  \cr
0 & 0 & 0 }$ 
\\
 & & & & &  \\
 \hline
 & & 1 & 3 (2) &-2& $\langle\Delta_1^L\rangle=\pmatrix{0,0,u_L}$
\\
\cline{3-6}
 Type-I see-saw mass &$\Delta_\zeta^L =
(\Delta_\zeta^{++},\Delta_\zeta^+,\Delta_\zeta^0)^L$  &$1'$&3 (2) &-2&
$\langle\Delta_2^L\rangle=\pmatrix{0,0,u_L}$
\\
\cline{3-6}
 & &$1''$&3 (2) &-2& $\langle\Delta_3^L\rangle=\pmatrix{0,0,u_L}$
\\
\hline
 & & & & &  \\
 Right-handed neutrino mass& $\hat{\Delta}^R=\pmatrix{
 \hat{\Delta}^0_1\cr
 \hat{\Delta}^0_2 \cr
 \hat{\Delta}^0_3 }^R
$&3&1 (0)&2&$\langle \hat{\Delta}^R \rangle= v_R \pmatrix{1 \cr
 \omega^2 \cr  \omega }$ 
\\
 & & & & &   \\
 \hline
 & & & &  & \\
 Right-handed neutrino mass
&$\Delta_3^R = (\Delta_3^{0})^R$  &$1''$&1 (0) &2&
$\langle\Delta_3^R\rangle={u}_R$ \\ 
& & & & &  \\
\hline
\end{tabular}
\end{center}
\caption{\em The scalar content of the model. The transformation
properties under $A4$ and $SU(2)_L$ are shown. The hypercharge
of the fields, $Y$, their lepton number, $L$, and the vacuum
expectation values are also indicated.} 
\label{tab1s}
\end{table}

All lepton masses arise from $A4$-invariant Yukawa-type
couplings. This requires several scalar fields\footnote{Alternate
$A4$ models address this issue by separating the $SU(2)_L$ and
$A4$ breakings \cite{A4af}.  The former proceeds through the
conventional doublet and triplet scalars which do not transform
under $A4$. The $A4$ breaking is triggered through the {\em vev}
of $SU(2)_L$ singlet `flavon' scalars which transform
non-trivially under $A4$. While economy is indeed a virtue here,
one pays a price in the form of the effective dimension-5
interactions which have to be introduced to couple the fermion
fields simultaneously to the two types of scalars.}  which develop
appropriate vacuum expectation values ({\em vev}).  To
generate the charged lepton masses one uses an $SU(2)_L$ doublet
$A4$ triplet of scalar fields $\Phi_i ~(i=1,2,3)$.   The Type-II
see-saw for left-handed neutrino masses requires $SU(2)_L$
triplet scalars.  The product rule in eq.  (\ref{A43x3})
indicates that these could be in the triplet
($\hat{\Delta}_i^L$), or the singlet 1, $1'$, $1''$
($\Delta_\zeta^L, ~\zeta=1,2,3$) representations of $A4$. As
discussed in the following, {\em all} of these are required to
obtain the dominant Type-II see-saw neutrino mass matrix of the
form of our choice. The Type-I see-saw results in a smaller
contribution whose effect is included perturbatively. For the
Dirac mass matrix of the neutrinos an $A4$ singlet $SU(2)_L$
doublet $\eta$, with lepton number -1, is introduced\footnote{
The assignment of opposite lepton numbers to $\nu_L$ and $N_R$
forbids their Yukawa coupling with $\Phi$ and the Dirac mass matrix
can be kept proportional to the identity.}.  The right-handed neutrino
mass matrix also arises from Yukawa couplings which respect $A4$
symmetry\footnote{Since the right-handed neutrinos are $SU(2)_L$
singlets, in principle, one can include direct Majorana mass
terms for them.  These dimension three terms would break $A4$
softly.}. The scalar fields required for this are all $SU(2)_L$
singlets and under $A4$ they transform as triplet
($\hat{\Delta}^R_i$) or the singlet  $1''$ ($\Delta^R_3$). The
scalar fields  of the model, their transformation properties
under the $A4$ and $SU(2)_L$ groups, their lepton numbers and
vacuum expectation values are summarized in Table \ref{tab1s}.

The Type-I and Type-II mass terms for the
neutrinos as well as the charged lepton mass matrix arise from
the $A4$ and $SU(2)_L$ conserving
Lagrangian\footnote{Note that the Dirac mass terms are also $L$ conserving.}:
\begin{eqnarray}
\mathscr{L}_{mass}&=& y_j \rho_{jik} \bar{l}_{Li} l_{Rj}\Phi_{k}^{0} 
 ~~{\rm(charged ~lepton ~mass)}  \nonumber\\
&+& f \rho_{1ik} \bar{\nu}_{Li}  N_{Rk} \eta^0 
 ~~{\rm(neutrino ~Dirac ~mass)} \nonumber\\
&+&\frac{1}{2}(\hat{Y}^L ~\alpha_{ijk}\nu_{Li}^TC^{-1}\nu_{Lj}
\hat{\Delta}_k^{L0}
+Y^L_{\zeta} ~\rho_{\zeta ij}\nu_{Li}^TC^{-1}\nu_{Lj}\Delta_\zeta^{L0})
 ~~{\rm(neutrino ~Type\!-\!II ~see\!-\!saw ~mass)} \nonumber \\
&+&  \frac{1}{2}(\hat{Y}^R ~\alpha_{ijk} N_{Ri}^TC^{-1}N_{Rj}
\hat{\Delta}_k^{R0}
+Y^R_{3} ~\rho_{3ij} N_{Ri}^TC^{-1} N_{Rj}\Delta_3^{R0}) 
~~{\rm(rh ~neutrino ~mass)} + h.c.  
\label{e1}
\end{eqnarray}

The scalar fields in the above Lagrangian get the following {\em
vev}s (suppressing the $SU(2)_L$ part):
\begin{equation}
\langle \Phi^0 \rangle = \frac{v}{\sqrt{3}} \pmatrix{1 \cr 1 \cr 1} \;,\;
\langle \hat{\Delta}^{L0} \rangle = v_L\pmatrix{1 \cr 0 \cr 0} \;,\; 
\langle \Delta_1^{L0} \rangle =  
\langle \Delta_2^{L0} \rangle =  
\langle \Delta_3^{L0} \rangle = {u}_L \;, 
\label{vev1}
\end{equation}
\begin{equation}
\langle \eta^0 \rangle = u  \;,\; 
\langle \hat{\Delta}^{R0} \rangle = 
v_R\pmatrix{1 \cr \omega^2 \cr \omega} \;,\; 
\langle \Delta_3^{R0} \rangle = {u}_R \;.\; 
\label{vev2}
\end{equation}
The scalar potential involving the fields listed in Table
\ref{tab1s} has many terms and is given in an Appendix. There we
indicate the conditions under which the scalars achieve the {\em
vev} listed in eqs. (\ref{vev1}) and (\ref{vev2}).

This results in  the charged
lepton mass matrix and the left-handed neutrino Majorana mass
matrix of the following forms:
\begin{equation}
M_{e\mu\tau}=\frac{v}{\sqrt{3}}
\pmatrix{y_1 & y_2 & y_3\cr
y_1 & \omega y_2 & \omega^2 y_3\cr
y_1 & \omega^2 y_2 & \omega y_3} \;\;,\;\; M_{\nu L}=
\pmatrix{(Y^L_1+ 2 Y^L_2)u_L & 0 & 0 \cr
0  & (Y^L_1 - Y^L_2)u_L & \hat{Y}^L v_L/2 \cr
0 & \hat{Y}^L  v_L/2 & (Y^L_1 - Y^L_2)u_L}\;\;.
\label{mmatrix1}
\end{equation}
where we have chosen $Y^L_2 = Y^L_3$. In the above the Yukawa
couplings satisfy $y_1 v = m_e, ~ y_2 v = m_\mu , ~ y_3 v =
m_\tau$.  The Type-II see-saw generates, $M_{\nu L}$, the
dominant contribution to the neutrino mass matrix. In the absence
of the solar splitting this involves just two masses $m_1^{(0)}$
and $m_3^{(0)}$. To obtain the requisite structure one must identify
$3 Y^L_1 u_L = 2[ 2 m_1^{(0)} -  m_3^{(0)}], ~3 Y^L_2 u_L = 
m_1^{(0)} + m_3^{(0)}$, and $\hat{Y}^L
v_L = 2[m_1^{(0)} + m_3^{(0)}]$.
The neutrino Dirac mass matrix and the mass matrix of the
right-handed neutrinos  are:
\begin{equation}
M_D = f u ~\mathbb{I}\;\;,\;\; 
M_{\nu R}= \pmatrix{Y^R_3 u_R & \hat{Y}^R v_R \omega/2  &
\hat{Y}^R v_R \omega^2/2 \cr \hat{Y}^R v_R
\omega/2  & Y^R_3 u_R \omega^2 & \hat{Y}^R v_R/2  \cr 
\hat{Y}^R v_R \omega^2/2 & \hat{Y}^R v_R/2  & Y^R_3
u_R \omega}\;\;.
\label{mmatrix2}
\end{equation}
The two unknown combinations appearing in $M_R$ above are
expressed as $Y^R_3 u_R \equiv
(2a + b)$ and $\hat{Y}^R v_R \equiv 2(b - a)$. 

The mass matrices in Eq. (\ref{mmatrix1}) can be put in a more
tractable form by using two transformations, the first being $U_L$ on the
left-handed fermion doublets and the other $V_R$ on the
right-handed neutrino singlets. $U_L$ and $V_R$ are given by  
\begin{equation}
U_L = {1 \over{\sqrt 3}}
\pmatrix{1 & 1 & 1 \cr
1 & \omega^2 &  \omega\cr
1 &  \omega&  \omega^2} = V_R \;.
\label{chngbasis}
\end{equation}
No transformation is applied on
the right-handed charged leptons. In the new basis, which we call
the {\em flavour} basis, the charged lepton mass matrix is
diagonal and the entire lepton mixing resides in the neutrino
sector. The mass matrices now are:
\begin{equation}
M_{e\mu\tau}^{flavour} = \pmatrix{m_e & 0 & 0 \cr 0 & m_\mu & 0 \cr
0 & 0 & m_\tau}\;\;,\;\;  
M_{\nu L}^{flavour} = {1 \over 2}
\pmatrix{2m^{(0)}_1 & 0 & 0 \cr 0 & m^+ & -m^- \cr 0 & -m^- & m^+} \;\;,
\label{mflav1}
\end{equation}
and 
\begin{equation}
M_D = f u ~\mathbb{I}\;\;,\;\;  
M_{\nu R}^{flavour} = {1 \over 2}
\pmatrix{0 & a & 0 \cr a & 0 & 0 \cr 0 & 0 & b} 
\;\;.
\label{mflav1a}
\end{equation}
Here $m^{\pm} = m_3^{(0)} \pm m_1^{(0)}$. $m^-$ is positive
for normal ordering (NO) of masses while it is negative for
inverted ordering (IO). We use the
notation $m_D = f u$.

\section{Model implications}

The $A4$ model we have presented results in the four mass
matrices in eqs. (\ref{mflav1}) and (\ref{mflav1a}).  The lepton
mixing and CP-violation will be determined, in this basis,
entirely by the neutrino sector on which we focus from here on.

The left-handed neutrino mass matrix $M_{\nu L}^{flavour}$,
obtained via a Type-II see-saw, dominates over the Type-I see-saw
contribution from the mass matrices in eq. (\ref{mflav1a}). The
contribution from the latter is included using perturbation
theory.

In the `mass basis' the left-handed neutrino mass matrix is
diagonal. The columns of the diagonalising matrix are the unperturbed
flavour eigenstates in this basis. We find from $M_{\nu L}^{flavour}$:
\begin{equation}
M^0 = M_{\nu L}^{mass} = U^{0T} M_{\nu L}^{flavour} U^0 = 
\pmatrix{m^{(0)}_1 & 0 & 0 \cr
0 & m^{(0)}_1 & 0 \cr 0 & 0 & m^{(0)}_3} \;\;,
\label{mass0}
\end{equation}
the orthogonal matrix, $U^0$, being
\begin{equation}
U^0=
\pmatrix{1 & 0 & 0 \cr
0 & {1\over\sqrt{2}} & -{1\over\sqrt{2}} \cr
0 & {1\over\sqrt{2}} & {1\over\sqrt{2}}}.
\label{mixing0}
\end{equation}
From eqs. (\ref{mass0}), (\ref{PMNS}) and (\ref{mixing0}) it is seen that the
solar splitting is absent, $\theta_{12} = 0, ~\theta_{13} = 0$,
$\delta = 0$, and $\theta_{23} = \pi/4$.

Before proceeding with the analysis we would like to remark on
the right-handed neutrino Majorana mass matrix in eq.
(\ref{mflav1a}), $M_{\nu R}^{flavour}$, which follows from the
$A4$ symmetric Lagrangian.  It has a four-zero texture. This
has the virtue of being of a form of $M_{\nu
R}^{flavour}$ with the most number of texture zeros. For the
see-saw to be operative, the matrix has to be invertible. This
eliminates matrices with five texture zeros in the flavour basis.
Of the 15 possibilities with four texture zeros there are only
two which are invertible and also meet the requirements of the
model (i.e., result in a non-zero $\theta_{12}$, $\theta_{13}$,
and shift $\theta_{23}$ from $\pi/4$).  These are:
\begin{equation}
M_1 =  {1 \over 2} \pmatrix{0 & a & 0 \cr a & 0 & 0 \cr 0 & 0 & b} \;\;, \;\;
M_2 =  {1 \over 2} \pmatrix{0 & 0 & a \cr 0 & b & 0 \cr a & 0 & 0} \;\;. \;\;
\label{texture4}
\end{equation}
Note, $M_1 \leftrightarrow M_2$ under $2 \leftrightarrow 3$
exchange\footnote{In the Lagrangian in eq. (\ref{e1}) the
replacement $\Delta_3^{R} \rightarrow \Delta_2^{R}$, where
$\Delta_2^{R}$ transforms like a $1'$ under  $A4$, yields an
$M_{\nu R}^{flavour}$ of the form of $M_2$.}. The results from
these two alternatives are very similar except for a few relative
signs in the interrelationships among $\theta_{13}$,
$\theta_{12}$, and $\theta_{23}$.  The $M_{\nu R}^{flavour}$
in eq. (\ref{mflav1a}) is of the form of $M_1$. We remark in the
end about the changes  which entail if the $M_2$ alternative is
used.

Taking $a$ and $b$ as complex we express $M_{\nu R}^{flavour}$ as:
\begin{equation}
 M_{\nu R}^{flavour} ={m_R} 
\pmatrix{0 & x e^{-i\phi_1} & 0 \cr x e^{-i\phi_1} & 0 & 0 \cr 0 & 0
& y e^{-i\phi_2}}  \  \ ,
\label{mflavgen}
\end{equation}
where $x, y$ are dimensionless real constants of ${\cal O}(1)$ and
$m_R$ sets the mass-scale. No
generality is lost by keeping the Dirac mass real.

In the flavour basis, the Type-I see-saw contribution, which we
treat as a perturbation, is:
\begin{equation}
 M'^{flavour} = \left[M_D^T(M_{\nu R}^{flavour})^{-1}M_D \right]  = 
{m_D^2 \over  xy m_R} 
\pmatrix{0 & y ~e^{i\phi_1} & 0 \cr y~e^{i\phi_1} & 0 & 0 \cr 
0 & 0 & x~e^{i\phi_2}} \;\;.
\label{pertmatcf}
\end{equation}
In the mass basis it is:
\begin{equation}
 M'^{mass} = U^{0T} M'^{flavour}U^0 
= {m_D^2 \over \sqrt 2 ~xy m_R} 
\pmatrix{0 & y~e^{i\phi_1} & -y~e^{i\phi_1} \cr 
y~e^{i\phi_1} & {x~e^{i\phi_2} \over  \sqrt 2} 
& {x~e^{i\phi_2} \over \sqrt 2} \cr 
-y~e^{i\phi_1} & {x~e^{i\phi_2} \over \sqrt 2} & 
{x~e^{i\phi_2} \over \sqrt 2}}\;\;.
\label{pertmatc}
\end{equation}

\section{Results}

After having presented the group-theoretic underpinnings of the
model we now indicate its predictions which could be tested in
the near future. As noted, from eq. (\ref{mixing0}) one has
$\theta_{23} = \pi/4$ and the other mixing angles are vanishing.
Further, once a choice of $m_0$, the lightest neutrino mass, is
made,  depending on the mass ordering either $m^{(0)}_1$ or
$m^{(0)}_3$ is determined. The remaining one of these two is
fixed so that the atmospheric mass splitting is correctly
reproduced. The solar mass splitting, $\theta_{12}$,
$\theta_{13}$ and the deviation of $\theta_{23}$ from maximality
are all realized through the first order perturbation,  which
results in inter-relationships between them. These offer a scope
of subjecting the model to experimental probing. From eq.
(\ref{mass0}) it is seen that to obtain the solar mixing
parameters one must take recourse to degenerate perturbation
theory.

\subsection{Data}
\label{sec:data}

From global fits the currently favoured 3$\sigma$ ranges of the neutrino
mixing parameters are
\cite{Gonzalez, Valle} 
\begin{eqnarray}
\Delta m_{21}^2 &=& (7.03 - 8.09) \times 10^{-5} \, {\rm eV}^2, \;\;
\theta_{12} = (31.30 - 35.90)^\circ, \nonumber \\
|\Delta m_{31}^2| &=& (2.325 -  2.599) \times 10^{-3}
\, {\rm eV}^2, \;\;
\theta_{23} = (38.4 - 53.3)^\circ \,, \nonumber \\
\theta_{13} &=& (7.87 - 9.11)^\circ, \;\; \delta =
(0 - 360)^\circ \;\;.
\label{results}
\end{eqnarray}
Here, $\Delta m_{ij}^2 = m_i^2 - m_j^2$, so that $\Delta
m_{31}^2 > 0 ~(<0)$ for normal (inverted) ordering. The data
indicate two best-fit points for $\theta_{23}$ in the first and
second octants. Later, we also use the recent preliminary T2K
hints \cite{Bronner} of $\delta$ being near -$\pi/2$.

\subsection{Real $M_R$ ($\phi_1 = 0
~{\rm or} ~\pi, \phi_2 = 0 ~{\rm or} ~\pi$)}\label{sec3}

$M_R$ is real if the phases $\phi_{1,2}$ in eq.  (\ref{mflavgen})
are 0 or $\pi$.  For notational simplicity, instead of retaining
these phases we allow $x$ ($y$) to be of either sign, thus
capturing the possibilities of $\phi_1$ ($\phi_2$) being 0 or
$\pi$.

In the real limit eq. (\ref{pertmatc}) becomes 
\begin{equation}
 M'^{mass} = 
{m_D^2 \over \sqrt 2 ~xy m_R} 
\pmatrix{0 & y & -y \cr y & {x \over  \sqrt 2} & {x \over \sqrt 2} \cr 
-y & {x \over \sqrt 2} & {x \over \sqrt 2}}\;\;.
\label{pert1}
\end{equation}
The effect of this perturbation on the degenerate solar sector is obtained
from the following $2\times2$ submatrix of the above,
\begin{equation}
M'^{mass}_{2\times2} = {m_D^2 \over \sqrt 2 ~xy m_R} 
\pmatrix{0 & y \cr y & {x/\sqrt 2}} \;.
\label{solr}
\end{equation}
This yields
\begin{equation}
\tan 2\theta_{12}= 2 \sqrt 2 \left(\frac{y}{x}\right)  \; .
\label{solangr}
\end{equation}
If  $y/x = 1$, i.e., $\hat{Y}^R = 0$ in eq. (\ref{mmatrix2}),
$\theta_{12}$ assumes the tribimaximal value,  which though
consistent at 3$\sigma$ is disallowed by the data at $1\sigma$.
From the data,  $\tan 2\theta_{12} > 0$ always, implying $x$ and
$y$ have to be either both positive or both negative.  In other words,
$\phi_1 = \phi_2$ and can be either 0 or $\pi$.  The fitted range
of $\theta_{12}$ translates to
\begin{equation}
0.682 < \frac{y}{x} < 1.075 ~{\rm at} ~3\sigma \;.
\label{t12lim}
\end{equation}
Eqn. (\ref{solr}) also implies
\begin{equation}
\Delta m^2_{solar}=  {m_D^2 \over xy m_R} ~m^{(0)}_1
\sqrt{x^2 + 8y^2}\;. 
\label{solspltr}
\end{equation} 
Including the first-order corrections from eq. (\ref{pert1}) the
wave function for the non-degenerate state, $|\psi_3\rangle$, becomes:
\begin{equation}
|\psi_3\rangle =
\pmatrix{-\kappa \cr -{1\over \sqrt 2}{(1 - \frac{\kappa}{\sqrt 2} {x\over y})} 
\cr {1\over \sqrt 2}{(1 + \frac{\kappa}{\sqrt 2} {x\over y})} } \ \ \ ,
\label{psi3_1}
\end{equation} 
with
\begin{equation}
\kappa \equiv {m_D^2 \over {\sqrt 2 ~x m_R m^-}} \;\;.
\label{kappa}
\end{equation} 
For $x > 0$  the sign of  $\kappa$ is the same as that
of $m^-$.
Comparing with the third column of eq. (\ref{PMNS})
one then has
\begin{equation}
\sin \theta_{13}\cos\delta=-\kappa= -{m_D^2 \over {\sqrt 2 ~x m_R m^-}}\;\;.
\label{s13}
\end{equation}

In the case of normal ordering $x > 0$ implies  $\delta = \pi$ while
for inverted  ordering $\delta = 0$, CP remaining 
conserved in both cases\footnote{The mixing angles
$\theta_{ij}$ are kept in the first quadrant and $\delta$ ranges
from $-\pi$ to $\pi$, as is the convention.}.
If $x < 0$ NO (IO) gives $\delta = 0 ~(\pi)$.  From
eqs. (\ref{s13}), (\ref{solangr}), and (\ref{solspltr}) one can write,
\begin{equation}
\Delta m^2_{solar} =  -{\rm sgn}(x)~m^- m^{(0)}_1 
~\frac{4 \sin \theta_{13} \cos\delta}{\sin 2\theta_{12}}  \;, 
\label{solsplr2}
\end{equation}
which relates\footnote{It is readily seen from eq.
(\ref{s13}) that $-sgn(x)m^- \cos\delta$ is always positive,
ensuring $\Delta m^2_{solar} > 0$.} the solar sector with
$\theta_{13}$.  Once the neutrino mass splittings, and the angles
$\theta_{12}$, and $\theta_{13}$ are given, eq.  (\ref{solsplr2})
fixes the lightest neutrino mass, $m_0$.

\begin{figure}[h]
\begin{center}
\includegraphics[width=3.2in]{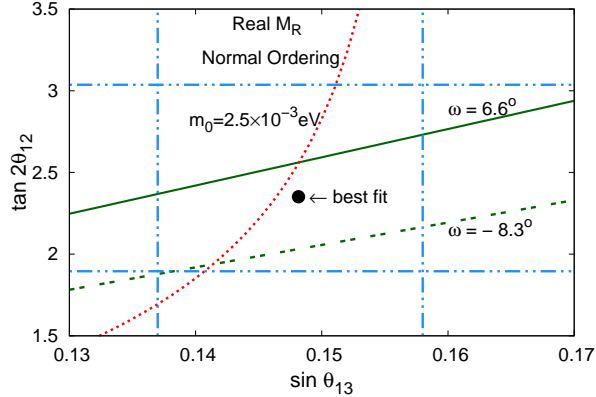}
\end{center}
\caption{ \em  The  area inside blue dot-dashed box  in the $\sin
\theta_{13}$ -  $\tan 2\theta_{12}$ plane is allowed by the
experimental  data at 3$\sigma$. The best-fit point is shown as a
black dot.  The red dotted curve gives the best-fit solar
splitting -- from eq. (\ref{solsplr2}) -- for $m_0 = 2.5$ meV.
Using eq. (\ref{phir}) for $\theta_{23}$ the area excluded at
3$\sigma$ is below the green solid (dashed) straight line for the
first (second) octant.  Only normal ordering is allowed for real
$M_R$. }
\label{Real1} 
\end{figure}


It can be checked that  eq. (\ref{solsplr2}) does not permit
inverted ordering. To this end, one defines 
$z \equiv m^- m^{(0)}_1/\Delta m^2_{atmos}$ and
$\tan \xi \equiv m_0/\sqrt{|\Delta m^2_{atmos}|}$. Note that $z$
is positive for both mass orderings and one has: 
\begin{eqnarray}
z &=& \sin \xi/(1+ \sin \xi) \;\; 
{\rm (normal ~ordering)},\nonumber \\
z &=& 1/(1+ \sin \xi) \;\; {\rm (inverted  ~ordering)} \;\;. 
\label{m_0}
\end{eqnarray} 
This implies $0 \leq z \leq 1/2$ for NO
while $1/2 \leq z \leq 1$ for IO. In both cases $z$ approaches 1/2 
as $m_0 \rightarrow $ large, i.e., one tends  towards quasi-degeneracy.
From eq. (\ref{solsplr2}) 
\begin{equation}
z = \left(\frac{\Delta m^2_{solar}}{|\Delta m^2_{atmos}|}\right) 
\left(\frac{\sin 2\theta_{12}}{4 \sin \theta_{13} |\cos\delta|}\right) \;\;.
\label{z}
\end{equation} 
Bearing in mind that  for real $M_R$ one has $|\cos\delta| = 1$
and using the measured values of the other oscillation parameters
one finds   $z\sim 10^{-2}$. This excludes the inverted mass
ordering option.

Further, eq. (\ref{psi3_1}) implies: 
\begin{equation}
\tan\theta_{23} \equiv \tan (\pi/4 - \omega) = 
\frac{1-\frac{\kappa}{\sqrt 2} {x\over y}}  
{1 + \frac{\kappa}{\sqrt 2} {x\over y}}   
\ \ . 
\label{th23r}
\end{equation}
Taken together with eqs. (\ref{solangr}) and (\ref{s13}) one has
\begin{equation}
\tan \omega = \frac{\kappa}{\sqrt 2} {x\over y} = 
-\frac{2 \sin \theta_{13}\cos\delta}{\tan2\theta_{12}} \;.
\label{phir}
\end{equation}
When $\omega$ is positive (negative), i.e., $\delta = \pi  ~(0)$,
we get $\theta_{23}$ in the
first (second) octant.  This corresponds to $x > 0$ ($x < 0$) for
NO, the allowed ordering for real $M_R$.

We are now in a position to state the consequences of this model
for real $M_R$. There are three independent input parameters, namely,
$m_0$, $\kappa$, and $y/x$ which determine
$\theta_{12}, \theta_{13}, \theta_{23}$, and  $\Delta
m^2_{solar}$ for NO.  For real $M_R$ inverted ordering is not
permitted.

In Fig. \ref{Real1} the main consequences of this
model  for real $M_R$ are displayed.  The region inside the blue
dot-dashed box is the 3$\sigma$ range of $\sin \theta_{13}$ and
$\tan 2\theta_{12}$ from the global fits, the best-fit point
being the black dot.  From the data in Sec. \ref{sec:data} for
both octants $\omega_{min} = 0$ at 3$\sigma$ and $\omega_{max} =
6.6^\circ ~(-8.3^\circ)$ for the first (second) octant. For the
$\omega_{max}$ for the first (second) octant eq. (\ref{phir}) of
this model corresponds to the green solid (dashed) straight
line, the area below being excluded.  Further, for real $M_R$,
as $|\cos\delta | = 1$, from eq.  (\ref{phir}) we get $|\omega|
\geq 5.14^\circ$ for both octants.  So far, we have not
considered the solar mass splitting.  Once $\Delta m^2_{solar}$
and $|\Delta m^2_{atmos}|$ are specified, the $z$ (or
equivalently $m_0$) that produces the correct solar splitting for
any chosen point in the plane is determined by eq.  (\ref{z}). In
this way, using the 3$\sigma$ ranges of $\theta_{13}$ and
$\theta_{12}$ one finds $z_{max} = 6.03 \times 10^{-2}$,
corresponding to   $(m_0)_{max}$ = 3.10 meV.  The consistency of
eq.  (\ref{z}) with eq.  (\ref{phir}) at $\omega_{max}$ sets
$z_{min}$ = 4.01 $\times 10^{-2}$ (3.88 $\times 10^{-2}$) for the
first (second) octant which translates to $(m_0)_{min}$ = 2.13
(2.06) meV.  As an example, choosing $m_0 = 2.5$ meV and taking
the best-fit points of the solar and atmospheric mass splittings
eq.  (\ref{solsplr2}) gives the red dotted curve in Fig.
\ref{Real1}.


\subsection{Complex $M_R$}
The shortcomings of the real $M_R$ case -- no CP-violation,
inverted ordering disallowed -- can be overcome when $M_R$ is
complex. One then has, as in eq. (\ref{pertmatc}),
\begin{equation}
 M'^{mass} =
{m_D^2 \over \sqrt 2 x y m_R} 
\pmatrix{0 & y e^{i\phi_1} & -y e^{i\phi_1} \cr y e^{i\phi_1} &
x {e^{i\phi_2}}\over{\sqrt{2}} & x {e^{i\phi_2}}\over{\sqrt{2}} \cr
-y e^{i\phi_1} & {x e^{i\phi_2}}\over{\sqrt{2}}&
x {e^{i\phi_2}}\over{\sqrt{2}}}.
\label{TypeIcmplxM}
\end{equation}

$x$ and $y$ are  now both positive. Since $M'$ 
is not hermitian any more 
the hermitian combination $(M^0 + M')^\dagger(M^0 + M')$
is considered, treating
$M^{0\dagger} M^0$ as the unperturbed piece and $(M^{0\dagger}
M' + M'^\dagger M^0)$ as the perturbation at lowest order. The
zero-order eigenvalues are $(m^{(0)}_i)^2$.  Written as a $3\times3$
hermitian matrix the perturbation  is
\begin{equation}
(M^{0\dagger} M' + M'^\dagger M^0)^{mass} = 
{m_D^2 \over \sqrt 2 xy m_R}
\pmatrix{ 0 & 2 m^{(0)}_1 y \cos\phi_1 & 
-y f(\phi_1) \cr
2 m^{(0)}_1 y \cos\phi_1 & { 2 \over \sqrt{2}}m^{(0)}_1 x \cos\phi_2 & 
{1\over\sqrt{2}} x f(\phi_2)\cr
-y f^*(\phi_1)& {1\over\sqrt{2}} x f^*(\phi_2)& 
{ 2 \over \sqrt{2}}m^{(0)}_3 x \cos\phi_2} \;.
\label{pertcmplx}
\end{equation}
Above
\begin{equation}
f(\varphi) = m^{+} \cos\varphi - i m^{-} \sin\varphi \;\;.
\label{ffn}
\end{equation}
Eqn. (\ref{pertcmplx}) provides the basis for the remaining calculation.

In a manner similar to the real $M_R$ case, from eq.
(\ref{pertcmplx}) we get
\begin{equation}
\tan 2\theta_{12}= 2\sqrt2 ~{y \over x}
~{\cos\phi_1\over\cos\phi_2} \;.
\label{solangcmplx}
\end{equation}
Since $\tan 2\theta_{12}$ remains positive at 3$\sigma$,
$\cos\phi_1$ and $\cos\phi_2$ must be of the same sign.  The
allowed possibilities for these phases are shown in Table
\ref{tabCP}. We can
take over the limits  in
eq. (\ref{t12lim}) which now apply on 
$(y/x)(\cos\phi_1/\cos\phi_2)$.

In place of eq. (\ref{psi3_1}) we now have at first order: 
\begin{equation}
|\psi_3\rangle =
\pmatrix{-\kappa f(\phi_1)/m^+\cr -{1\over \sqrt 2}{[1-{\kappa \over \sqrt
2}\frac{x}{y} ~f(\phi_2)/m^+]} \cr {1\over \sqrt 2}{[1+{\kappa\over \sqrt
2}  \frac{x}{y} ~f(\phi_2)/m^+]}
} .
\label{psi3ca}
\end{equation}
Since $x,y$ are now positive quantities, the sign of $\kappa$ is
determined by that of $m^-$, i.e., $\kappa$ is positive
(negative) for normal  (inverted) ordering.
From eqs. (\ref{PMNS}) and  (\ref{psi3ca}) 
\begin{eqnarray}
\sin \theta_{13}\cos\delta &=& -\kappa  \cos\phi_1 \ , 
\nonumber \\ 
\sin \theta_{13}\sin\delta &=& -\kappa ~\frac{m^-}{m^+} \sin\phi_1\  \ . 
\label{s13cmplx}
\end{eqnarray}
Using eq.  (\ref{s13cmplx}) one can  immediately relate the
quadrant of $\delta$ with that of $\phi_1$ for both orderings.
These are  also presented in Table \ref{tabCP}. It can be
seen that a near-maximal $\delta = -\pi/2 - \epsilon$ is obtained
for normal (inverted) ordering if $\phi_1 \sim$ -$\pi/2 -
\epsilon$ (-$\pi/2 +
\epsilon$).

\begin{table}[tbh]
\begin{center}
\begin{tabular}{|c|c|c|c|c|c|}
\hline
$\phi_1$& $\phi_2$ &\multicolumn{2}{|c|}{Normal Ordering}  & 
\multicolumn{2}{c|}{Inverted Ordering} \\ \cline{3-6} 
quadrant& quadrant & $\delta$ & $\theta_{23}$& $\delta$ & $\theta_{23}$  \\ 
 & & quadrant & octant & quadrant & octant \\ \hline
$0  -  \pi/2$ & $0  -  \pi/2$ or -$\pi/2 ~-$  0 & -$\pi ~-$  -$\pi/2$ &
$0  -  \pi/4$& -$\pi/2 ~-$  0 &$\pi/4 - \pi/2$    \\ \hline
$\pi/2  -  \pi$ & $\pi/2  -  \pi$ or  -$\pi ~-$  -$\pi/2$ &  -$\pi/2 ~-$  0  &
$\pi/4  -  \pi/2$  & -$\pi ~-$  -$\pi/2$ & $0  -  \pi/4$  \\ \hline
 -$\pi ~-$  -$\pi/2$ & $\pi/2  -  \pi$ or  -$\pi ~-$  -$\pi/2$ & $0  -  \pi/2$ &
$\pi/4  -  \pi/2$ & $\pi/2  -  \pi$ &$0  -  \pi/4$    \\ \hline
-$\pi/2 ~-$  0 & $0  -  \pi/2$ or -$\pi/2 ~-$  0 & $\pi/2  -  \pi$ &
$0  -  \pi/4$ & $0  -  \pi/2$ & $\pi/4 - \pi/2$ \\
\hline
\end{tabular}
\end{center}
\caption{\em The options for the phase $\phi_1$ in $M_R$ and the
consequent ranges of the other phase $\phi_2$ in $M_R$, the
leptonic CP-phase $\delta$, and the octant of $\theta_{23}$ for
both mass orderings. All angles are in radians.  For inverted
ordering or quasi-degeneracy $\delta \sim \pi/2$ or $-\pi/2$. }
\label{tabCP}
\end{table}

In addition, for $\theta_{23}$ eq. (\ref{psi3ca}) implies 
\begin{equation}
\tan\theta_{23} = {{1-{\kappa \over
\sqrt 2}\frac{x}{y}\cos\phi_2} \over  {1+{\kappa\over \sqrt
2}\frac{x}{y} \cos\phi_2}}  \ \ \ \ .
\label{phic1}
\end{equation}
The deviation from maximality, $\omega$, can be obtained from the
above and using  eqs. (\ref{solangcmplx}) and (\ref{s13cmplx})
expressed as 
\begin{equation}
\tan\omega 
= -\frac{2\sin \theta_{13}\cos\delta}{\tan2\theta_{12}} \;,
\label{phic}
\end{equation}
which has the same form as eq. (\ref{phir}) for the real $M_R$
case except that now $\cos\delta$ can deviate from $\pm$1. 
The octant of $\theta_{23}$ for different choices of $\phi_1$
quadrants is  given in Table \ref{tabCP} for both mass
orderings.

Substituting for $m_D^2/m_R$ in terms of $\sin
\theta_{13}\cos\delta$, using eq. (\ref{s13cmplx}) one has from
eq. (\ref{pertcmplx})
\begin{equation}
\Delta m^2_{solar}
= - ~{\rm sgn}(\cos\phi_2) ~m^- m^{(0)}_1 
~\frac{4 \sin \theta_{13} \cos\delta }{\sin 2\theta_{12}} 
\;,
\label{solsplc}
\end{equation}
which is reminiscent of  eq. (\ref{solsplr2}) for real $M_R$.
Keeping in mind that $\cos\phi_1/\cos\phi_2$ must be positive and
using eq. (\ref{s13cmplx}) it is easy to see that $\Delta
m^2_{solar} > 0$ always. Since  the sign of $\omega$ -- i.e., the
octant of $\theta_{23}$ -- depends on the quadrant of only
$\cos\phi_1$, irrespective of the mass ordering one can
accommodate both octants while meeting the solar splitting requirement.

As in eq. (\ref{z}) one again has
\begin{equation}
|\cos\delta| = \left(\frac{\Delta m^2_{solar}}{|\Delta m^2_{atmos}|}\right) 
\left(\frac{\sin 2\theta_{12}}{4 \sin \theta_{13} ~z}\right) \;\;,
\label{cdel}
\end{equation} 
with the further proviso that $|\cos\delta|$ can now be anywhere
between zero and unity. This freedom removes the bar which
applied on inverted ordering for real $M_R$.

\begin{figure}[h]%
\begin{center}
{\includegraphics[scale=0.65,angle=0]{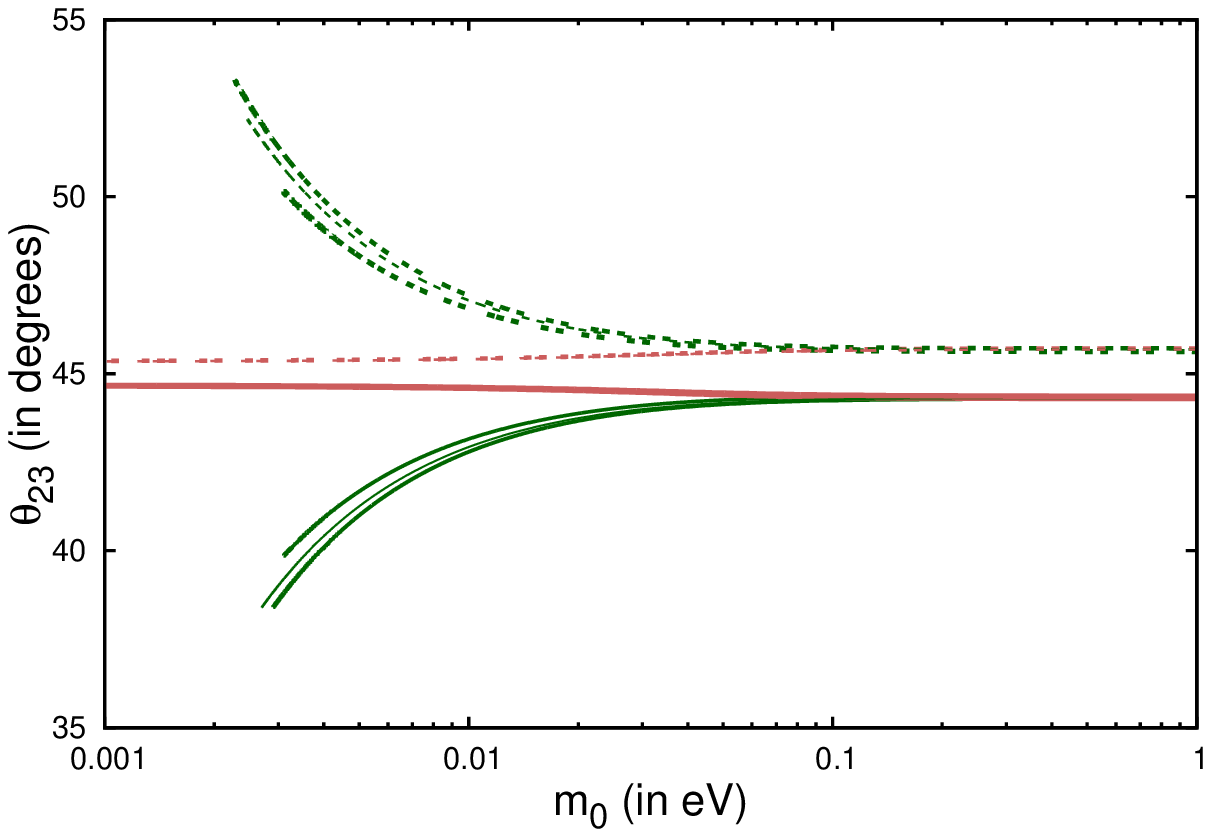}
\hspace*{0.25pt}
\includegraphics[scale=0.65,angle=0]{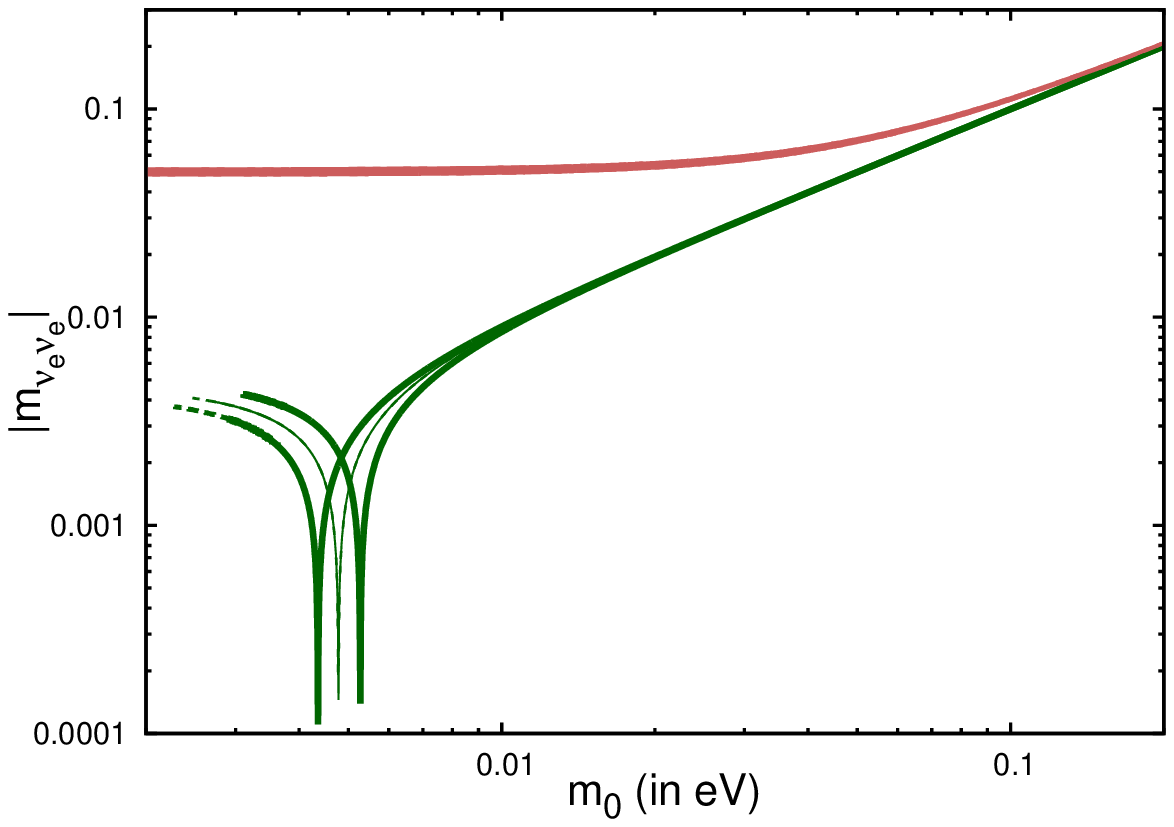}}
\caption{\em   $\theta_{23}$ (left panel) and $|m_{\nu_e
\nu_e}|$, the quantity controlling neutrino-less double-beta
decay, in eV (right panel) as a function of the lightest
neutrino mass $m_0$ (in  eV).   The green
(pink) curves are for NO (IO). The 3$\sigma$ allowed region is
between the thick curves while the thin curves are
for the best-fit input values. The solid (dashed) curves are for
the first (second) octant of $\theta_{23}$.    }
\label{th23} 
\end{center} 
\end{figure} 

\begin{figure}[h]%
\begin{center}
{\includegraphics[scale=0.65,angle=0]{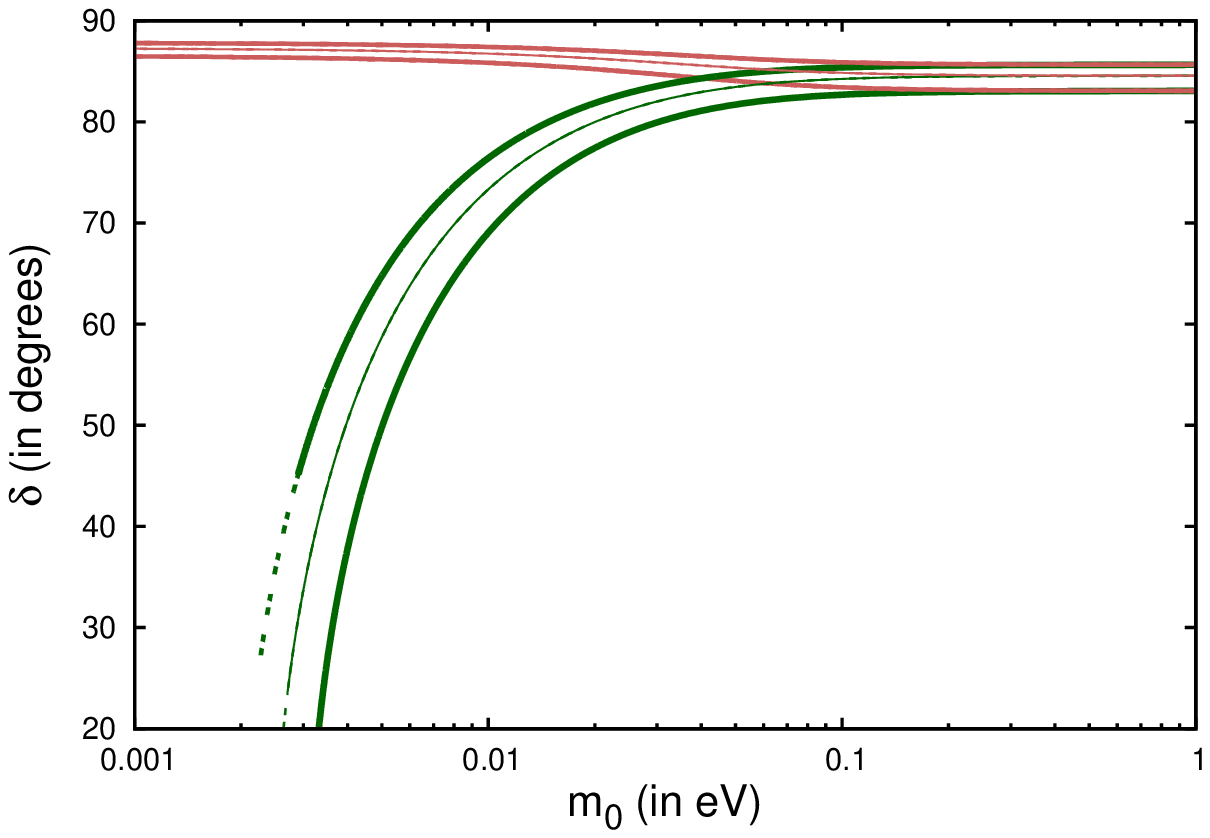}
\hspace*{0.25pt}
\includegraphics[scale=0.65,angle=0]{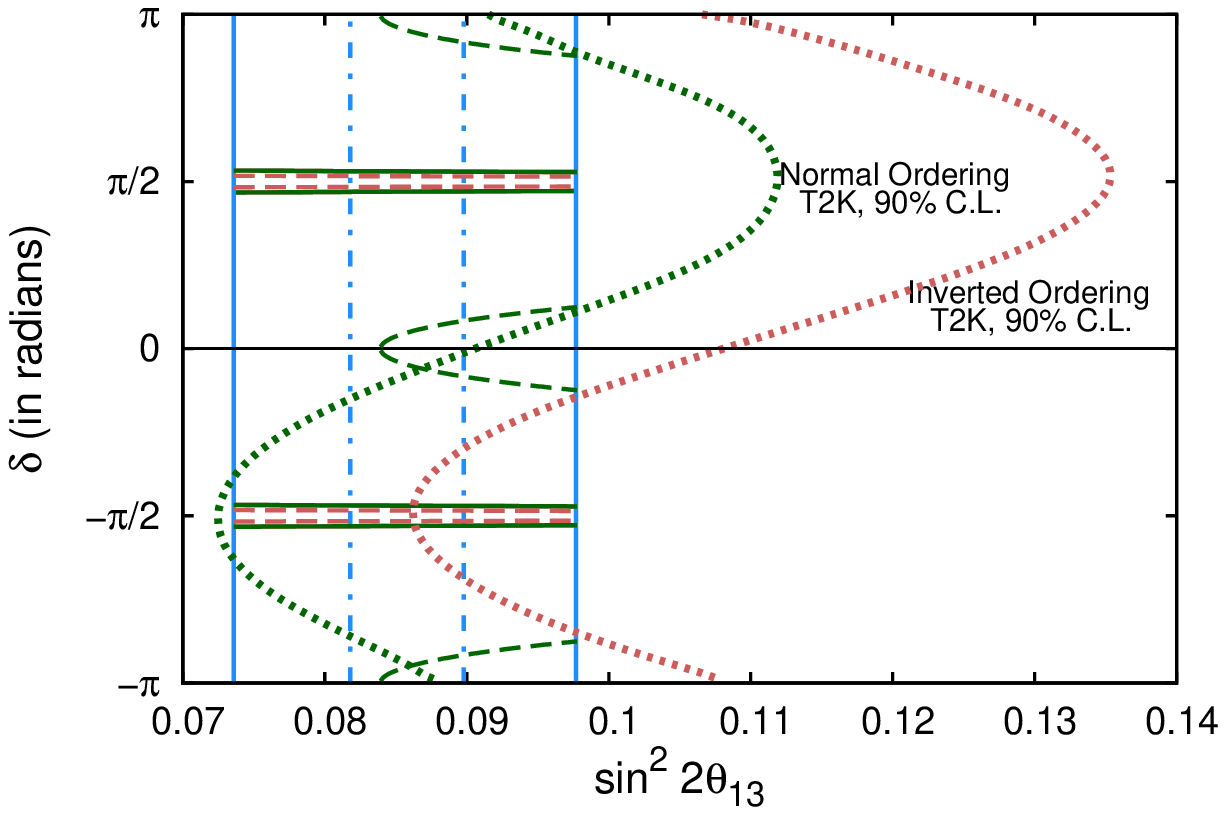}}
\caption{\em   The CP-phase $\delta$ from this model.   The green
(pink) curves are for NO (IO). Left: $\delta$ as a function of
$m_0$ in eV. The line-type conventions are as in Fig. \ref{th23}.
Results are shown only for the first quadrant. Right: The blue
vertical solid (dot-dashed) lines are the 3$\sigma$ (1$\sigma$)
allowed ranges of $\sin ^22\theta_{13}$ from global fits.
Dependence of $\delta$ for $m_0 = 0.5$ eV ($m_0 = 2.5$ meV) on
$\sin ^22\theta_{13}$ within the allowed range are the solid
(dashed) lines.  The curves for $m_0 = 0.5$ eV for the two
orderings overlap.  Also shown are the 90\% C.L. curves (dotted)
obtained by T2K \cite{Bronner} which disallow the region to their
left.  }
\label{CP} 
\end{center} 
\end{figure}

Here we use $m_0,  \theta_{13}$, and $\theta_{12}$ as inputs to
fix the model parameters.  Eqs. (\ref{phic}) and (\ref{cdel})
then determine $\theta_{23}$ and  $\delta$ respectively, as shown
in Figs. \ref{th23} and \ref{CP}. One can also obtain
$|m_{\nu_e \nu_e}|$, which determines the neutrino-less
double-beta decay rate, in terms of the mass eigenvalues and the
mixing angles. In these figures the green (pink) curves are for
NO (IO).

In the left panel of Fig. \ref{th23} the dependence of
$\theta_{23}$ on $m_0$ is presented while the right panel
shows $|m_{\nu_e \nu_e}|$ again as a function of $m_0$. The
thick lines delimit the 3$\sigma$ allowed regions while the thin
lines correspond to the best-fit values of input parameters. The
solid (dashed) curves are for the first (second) octant of
$\theta_{23}$.  The thick and thin curves for IO overlap and
cannot be distinguished.  As expected from eq.  (\ref{phic})
$\theta_{23}$ is symmetric about $\pi/4$. The experimental
3$\sigma$ bounds on $\theta_{23}$ for both octants determine a
minimum permitted $m_0$ for NO.  For IO there is no such lower
bound.  Planned experiments to measure the neutrino mass
\cite{katrin} are sensitive to $m_0$ not less than 200 meV.  From
Fig. \ref{th23} it is seen that at such a scale the two mass
orderings have close predictions, which is a reflection of
quasi-degeneracy.

In  the left panel of Fig. \ref{CP} we show the dependence of
$\delta$ on $m_0$  for both NO and IO. The line-type conventions
are the same as in Fig. \ref{th23}.  As noted in Table
\ref{tabCP} and eq. (\ref{s13cmplx}), $\delta$ can be in any of
the four quadrants depending on the choice of $\phi_1$.  Eq.
(\ref{cdel}) indicates that for all these four cases, namely,
$\pm \delta$ and $\pm(\pi - \delta)$, the dependence of
$|\cos\delta|$ on $m_0$ is identical for a chosen mass ordering.
Keeping this in mind, Fig. \ref{CP} (left panel)  has been
plotted with $\delta$ in the first quadrant.  For $m_0$ smaller
than $\sim$ 10 meV, $\delta$ differs significantly for the two
orderings.  As expected from Fig. \ref{Real1}, the real $M_R$
limit, i.e., $\delta = 0$, is obtained only for NO.

The variation of $\delta$ with  $\sin ^22\theta_{13}$ obtained
from eq.  (\ref{solsplc}) for both mass orderings for two
representative values of $m_0$ = 0.5 eV  (solid curves) and 2.5
meV  (dashed curves) is shown in the right panel of Fig.
\ref{CP}. Here the best-fit values of the two mass splittings and
$\theta_{12}$ have been used.  The allowed range of   $\sin
^22\theta_{13}$ from the global fits at 3$\sigma$ (1$\sigma$) is
bounded by the blue solid (dot-dashed) vertical lines. Note that in
all cases there are solutions for $\delta$ in every quadrant.
For IO $\delta$ remains close to $\pm\pi/2$ for all $m_0$. For
NO, with $m_0$ = 0.5 eV, which is in the quasi-degenerate region,
$\delta$ is the same as for IO while for $m_0$ = 2.5 meV one
finds $\delta$ around 0 or $\pm \pi$ and that too for a limited
range of $\sin ^22\theta_{13}$. In this panel we have also shown
90\% C.L. exclusion limits in the $\sin ^22\theta_{13} - \delta$ plane
-- dotted curves -- identified by the T2K collaboration. The
regions to the left of the curves are disfavoured. Notice that
$\delta = -\pi/2$ is preferred, which in our model is consistent
with IO for all masses but a limited range of $\sin
^22\theta_{13}$ while for NO though the full range of the latter
is consistent one must have $m_0 \geq$ 100 meV.  More precise
measurements of neutrino parameters will test this model
closely.

Finally, it is noteworthy that the phase $\phi_2$ enters
only in three places: in the combination $x \cos \phi_2/y$ in
the expressions for $\tan 2\theta_{12}$ and $\tan \theta_{23}$  -
eqs. (\ref{solangcmplx}) and (\ref{phic1}),
and as ${\rm sgn}(\cos\phi_2)$ in the formula for the solar
splitting -- eq. (\ref{solsplc}). So, its effect can be entirely
subsumed by redefining $\cos \phi_2/y \rightarrow 1/y$ and
permitting $y$ to be both positive and negative. Therefore for
complex $M_R$ the free input parameters are really
$m_0$, $\kappa$, $y/x$ and $\phi_1$ which determine the three
mixing angles, the solar mass splitting, and the CP-phase $\delta$.

Before concluding, we would like to make a comment on our choice of
$M_R$. In eq. (\ref{texture4}) two four-zero textures, $M_1$ and
$M_2$, were identified both of which could be admissible for
$M_R$. We had chosen $M_1$ for this work. If instead, we had
chosen $M_2$ then the discussion would go through essentially
unchanged except for  the replacement $\kappa
\rightarrow -\kappa$.


\section{Conclusions}

We have proposed a model for neutrino masses and lepton mixing
which relies on an underlying $A4$ symmetry.  All masses are
generated from  $A4$ invariant Yukawa couplings.  There are
contributions to the neutrino masses from both Type-I and Type-II
see-saw terms of which the latter is dominant.  It generates the
atmospheric mass splitting and keeps the mixing angles either
maximal, e.g., $\pi/4$ for $\theta_{23}$, or vanishing, for
$\theta_{13}$ and $\theta_{12}$. The solar splitting is absent.
The Type-I see-saw contribution, which is treated perturbatively,
results in  $\theta_{13}$, $\theta_{12}$, and $\theta_{23}$
consistent with the global fits and generates the solar
splitting. Both octants of $\theta_{23}$ are permitted.  Testable
relationships between these quantities, characteristic of this
model,  are derived. As another example, inverted ordering of
neutrino masses is correlated with a near-maximal CP-phase
$\delta$ and allows arbitrarily small neutrino masses. For normal
ordering $\delta$ can vary over the entire range and approaches
maximality in the quasi-degenerate limit. The lightest neutrino
mass cannot be lower than a few meV in this case.

While this paper was being finalised, NO$\nu$A announced
\cite{NOVA} their preliminary results based on the equivalent of
2.74 $\times 10^{20}$ p.o.t.  With $\sin^2\theta_{23} = 0.50$
the data favour NO and at 90\% CL indicate $\delta$ between
$-\pi$ to 0 with a preference for $\delta \sim -\pi/2$. As seen
from Fig. \ref{CP}  this is consistent with our model, with
$\delta \sim -\pi/2$ favouring $m_0$ in the quasi-degenerate
regime, i.e., $m_0 \geq \mathcal{O}$(0.1 eV).  If this result
is confirmed by further analysis then the model will require
neutrino masses to be  in a range to which ongoing
experiments will be sensitive \cite{numass}.

{\bf Acknowledgements:} SP acknowledges support from CSIR, India.
AR is partially funded by  the Department of Science and
Technology Grant No. SR/S2/JCB-14/2009.

\section*{Appendix: The scalar potential minimization}

\setcounter{equation}{0}  
\setcounter{section}{1}  

\renewcommand{\thesection}{\Alph{section}}
\renewcommand{\theequation}{\thesection.\arabic{equation}}  

In this Appendix we discuss the nature of the scalar
potential of the model in some detail. We also identify the
conditions which must be satisfied by the parameters of the
potential so that the {\em vev}s take the values
considered in the model. Needless to say, the conditions ensure
that the potential is locally minimized by this choice. To check
whether it is also a {\em global} minimum is beyond the scope of
this paper\footnote{For example, the global minima of the
relatively simple case of one $A4$ triplet $SU(2)_L$ doublet
scalar multiplet have been identified in \cite{gmin1} and used in
the context of a model for leptons in \cite{gmin2}.}.

The scalars listed in Table 2 have fields transforming under $A4$,
$SU(2)_L$, and $U(1)_Y$ which also carry a lepton number. The
scalar potential has to be of the most general quartic form which
is a singlet under all these symmetries. Below we include all
terms that are permitted by the symmetries. Invariance under
$SU(2)_L$, $U(1)_Y$ and lepton number are easy to verify.

\subsection{$A4$ invariants: Notation and generalities}

Here we give a brief account of our notation and the $A4$-invariant terms.
First recall that there are scalars which transform as $1, 1',
1'',$ and $3$ under $A4$. One must include in the potential up to
quartics in these fields which give rise to $A4$ singlets.  The
product rules of the one-dimensional representations $1, 1'$ and
$1''$ are simple, it is the $A4$ triplet which requires some
discussion. To this end consider two $A4$ triplet fields
$A\equiv(a_1,a_2,a_3)^T$ and $B\equiv(b_1,b_2,b_3)^T$ where $a_i,
b_i$ may have $SU(2)_L \times U(1)_Y$ transformation properties
which we suppress here.  As noted in eq. (\ref{A43x3}),
combining $A$ and $B$ one can obtain
\begin{equation}
3_A\otimes3_B = 1 \oplus 1' \oplus 1'' \oplus 3 \oplus 3  \;\;.
\label{A1}
\end{equation}

We denote the irreducible representations on the right-hand-side
by  $O_1(A, B)$, $O_{2}(A, B)$, $O_{3}(A, B)$, $T_s(A, B)$ and
$T_a(A, B)$, respectively, where, as noted in eqs.
(\ref{3x3to3}, \ref{3x3to1})
\begin{eqnarray}
O_{1}(A, B)&\equiv& 1 = a_1 b_1+a_2 b_2+a_3 b_3 \equiv
\rho_{1ij}a_i b_j \;\;,\nonumber \\
O_{2}(A, B)&\equiv& 1'= a_1 b_1+\omega^2 a_2 b_2+\omega a_3 b_3\equiv
\rho_{3ij}a_i b_j \;\;,  \nonumber \\
O_{3}(A, B) &\equiv& 1''= a_1 b_1+\omega a_2 b_2+\omega^2 a_3 b_3
\equiv \rho_{2ij}a_i b_j
\;\;,
\label{A2}
\end{eqnarray}  
and
\begin{eqnarray}
T_s(A, B)&\equiv& 3= \left(\frac{a_2 b_3 + a_3 b_2}{2}, \ \
\frac{a_3 b_1 + a_1 b_3}{2}, \ \
\frac{a_1 b_2 + a_2 b_1}{2}\right)^T
  \;\;,\nonumber\\
T_a(A, B)&\equiv& 3= \left(\frac{a_2 b_3 - a_3 b_2}{2}, \ \ \frac{a_3 b_1 - a_1 b_3}{2}, \ \ 
\frac{a_1 b_2 - a_2 b_1}{2}\right)^T \;\;.
\label{A3}
\end{eqnarray}
Note that $O_3(A^\dagger,A) = [O_2(A^\dagger,A)]^\dagger$ and
$T_a(A, A) = 0$.

The scalar potential can be written in this notation keeping in
mind the following:
\begin{itemize}
\item No two scalar multiplets have all quantum numbers  the
same.  So terms in the potential cannot be related by replacing
any one field by some other. 
\item Neither is there a scalar which is a singlet  under all the
symmetries. 
\end{itemize}
Thus the scalar potential can consist of terms of the
following forms (displaying only $A4$ behaviour):
\begin{enumerate} 
\item Quadratic: $W^\dagger W$,
\item Cubic: $X_{i} X'_{j} X''_{k}, X_{i} X_{j} X_{k}$,
$X'_{i} X'_{j} X'_{k},$ $X''_{i} X''_{j} X''_{k}$,
$O_{1}(Y_i,Y_j) X_k$, $O_{2}(Y_i,Y_j) X''_k$, 
$O_{3}(Y_i,Y_j) X'_k$,
\item Quartic: $(W_i^\dagger W_i)(W_j^\dagger W_j), ~(X_i X_j)
(X_k X_l), ~(X_i X_j) (X'_k X''_l), ~(X'_i X''_j)
(X'_k X''_l), ~(X'_i X'_j)(X'_k X_l), ~(X''_i X''_j)(X''_k X_l)$,\\
$O_{1}(Y_i,Y_j) X_k X_l, ~O_{1}(Y_i,Y_j) X'_k X''_l,
~O_{2}(Y_i,Y_j) X'_k X'_l, ~O_{2}(Y_i,Y_j) X_k X''_l$,\\
$O_{3}(Y_i,Y_j) X''_k X''_l, ~O_{3}(Y_i,Y_j) X_k X'_l$,\\
$O_{1}(Y_i,Y_j) O_{1}(Y_k,Y_l), O_{2}(Y_i,Y_j)^\dagger
O_{2}(Y_k,Y_l), O_{3}(Y_i,Y_j)^\dagger O_{3}(Y_k,Y_l),
O_{2}(Y_i,Y_j) O_{3}(Y_k,Y_l)$,\\
 $O_{1}(T_s(Y_i,Y_j), T_s(Y_k,Y_l))$,
$O_{1}(T_s(Y_i,Y_j), T_a(Y_k,Y_l))$,
$O_{1}(T_a(Y_i,Y_j), T_a(Y_k,Y_l))$.\\
$O_1(T_s(Y_i,Y_j), Y_k) X_l$, $O_{2}(T_s(Y_i,Y_j), Y_k) X''_l$,
$O_{3}(T_s(Y_i,Y_j), Y_k) X'_l$, \\
$O_1(T_a(Y_i,Y_j), Y_k) X_l$, $O_{2}(T_a(Y_i,Y_j), Y_k) X''_l$,
$O_{3}(T_a(Y_i,Y_j), Y_k) X'_l$ .
\end{enumerate}

In the above $W$ is any field, $X$, $X'$, and $X''$ stand for
generic fields transforming as $1$, $1'$, and $1''$ under $A4$
while $Y$ is a generic $A4$ triplet field.  We have not
separately listed the invariants formed using $X^{\dagger}$,
$X'^{\dagger}$, $X''^{\dagger}$, and $Y^{\dagger}$. 

Because of the large number of scalar fields in our model --
e.g., $SU(2)_L$ singlets, doublets, and triplets -- the scalar
potential has many terms.  To simplify this discussion, we exclude
cubic terms in the fields and take all couplings in the
potential to be real. For ease of presentation, we list the
potential in separate pieces:  (a) those restricted to any one
$SU(2)_L$ sector, and (b) those coupling scalars of different $SU(2)_L$
sectors. Since the {\em vev} of the $SU(2)_L$ singlets, which are
responsible for the right-handed neutrino mass, are much
larger than that of the other scalars, in the latter category we
keep only the terms which couple the singlet fields to either the
doublet or the triplet sectors.

\subsection{$SU(2)_L$ Singlet Sector:}
In the $SU(2)_L$ singlet scalar sector there is an $A4$ triplet
$\hat \Delta^R$  and another scalar $\Delta_3^R$ that transforms
as a $1''$. Eq. (\ref{A1}) shows that two
$\hat{\Delta}^{R}$ triplets can combine to give different $A4$
irreducible representations. For this purpose we introduce the
notations:
\begin{equation}
O_{1}^{ss} \equiv O_{1}(\hat{\Delta}^{R\dagger}, \hat\Delta^R);
~O_{2}^{ss} \equiv O_{2}(\hat{\Delta}^{R\dagger}, \hat\Delta^R);
~T_s^{ss} \equiv T_s(\hat{\Delta}^{R}, \hat\Delta^R).
\label{Prod3s}
\end{equation}
Generically, we will use the notation $\widetilde{O}_i$ or
$\widetilde{T}_{s,a}$ if the second $A4$ triplet field in the
argument is
replaced by its hermitian conjugate.  For example, here 
\begin{equation}
~\widetilde{O}_{3}^{ss} \equiv O_{3}(\hat{\Delta}^{R\dagger},
\hat\Delta^{R\dagger}) ~{\rm and}
~\widetilde{T}_s^{ss} \equiv T_s(\hat{\Delta}^{R}, \hat\Delta^{R\dagger}).
\end{equation}
 We will also require the combinations:
\begin{equation}
~\mathscr{O}_{2}^{ss} \equiv O_{2}(\hat{\Delta}^{R},
{T}_s^{ss\dagger}).
\end{equation}
From the $A4$ singlet $\Delta_3^R$ one can make the combination 
\begin{equation}
Q_{3}^{ss} \equiv \Delta_3^{R\dagger} \Delta_3^R \;\;,
\end{equation}
which is obviously a singlet under all the symmetries.

Using this notation the most general
scalar potential of this sector is given by:
\begin{eqnarray}
V_{singlet}&=& m_{\Delta_3^R}^2 Q_3^{ss} + m_{\hat\Delta^R}^2 O_{1}^{ss} 
+ {1 \over 2} \lambda^s_1\left[Q_3^{ss}\right]^2 
+{1 \over 2} \lambda^s_2 \left\{ [O_{1}^{ss}]^2 +
 (O_{2}^{ss})^\dagger O_{2}^{ss} 
+ O_{1}({T_s^{ss}}, T_s^{ss\dagger})  \right  \}
\nonumber \\ 
&+& 
{1 \over 2} \lambda^s_3\left[Q_3^{ss} O_{1}^{ss} \right]
+ \lambda^s_4 \left[ \mathscr{O}_{2}^{ss} \Delta_3^R + ~{\rm h.c.}\right]  
+ \lambda^s_5 \left[ \widetilde{O}_{3}^{ss} \Delta_3^{R} \Delta_3^{R}
 + ~{\rm h.c.}\right]  \;\;.
\label{Vs}
\end{eqnarray}
In the above, we have taken $\lambda^s_2$ as the common
coefficient of the different A4-singlets that can be obtained
from the combination of two $\hat\Delta^R$ and two $(\hat{\Delta}^R)^\dagger$
fields. We also follow a similar principle for the fields with other
$SU(2)_L$ behaviour.

\subsection{$SU(2)_L$ Doublet Sector:}

The $SU(2)_L$ doublet scalar sector comprises of the two fields 
$\Phi$ and $\eta$ transforming as $3$ and $1$ under $A4$ 
respectively. Recall that $\Phi$ and $\eta$ have opposite
hypercharge. In analogy to the singlet sector we denote the
required $A4$ triplet $\Phi$ combinations as:
\begin{equation}
O_{1}^{dd} \equiv O_{1}(\Phi^\dagger,\Phi);
~O_{2}^{dd} \equiv O_{2}(\Phi^\dagger,\Phi);
~T_s^{dd} \equiv T_s(\Phi,\Phi),
\end{equation}
and from the $A4$ singlet $\eta$
\begin{equation}
Q_{\eta}^{dd} \equiv \eta^{\dagger} \eta\;\;.
\end{equation}

The potential for this sector is:
\begin{eqnarray}
V_{doublet}&=& m_\eta^2 Q_{\eta}^{dd}  + m_\Phi^2 O_{1}^{dd} 
+ {1 \over 2} \lambda^d_1\left[Q_{\eta}^{dd}\right]^2
+{1 \over 2} \lambda^d_2 \left\{[O_{1}^{dd}]^2 + 
\{O_{2}^{dd}\}^\dagger O_{2}^{dd} \right.
\nonumber \\ 
&+& \left.
O_{1}({T_s^{dd}}, T_s^{dd\dagger})  \right \}
+{1 \over 2} \lambda^d_3\left[Q_{\eta}^{dd} O_{1}^{dd}\right].
\label{Vd}
\end{eqnarray}

\subsection{$SU(2)_L$ Triplet Sector:}

\vskip 10pt
The $SU(2)_L$ triplet sector consists of four fields, viz, 
$\hat{\Delta}^L$, $\Delta^L_1$, $\Delta^L_2$ and $\Delta^L_3$
transforming as $3$, $1$, $1'$, $1''$ under $A4$.
\vskip 1pt
We define
\begin{equation}
O_{1}^{tt} \equiv O_{1}(\hat{\Delta}^{L\dagger}, \hat\Delta^L);
~O_{2}^{tt} \equiv O_{2}(\hat{\Delta}^{L\dagger}, \hat\Delta^L);
~T_s^{tt} \equiv T_s(\hat{\Delta}^{L}, \hat\Delta^L),
\end{equation}
\begin{equation}
Q_{i}^{tt} \equiv \Delta_i^{L\dagger} \Delta_i^L \;\;, \;\; (i=1,2,3),
\end{equation}
and
\begin{equation}
~\mathscr{O}_{i}^{tt} \equiv O_{i}(\hat{\Delta}^{L},
{T}_s^{tt\dagger}) \;\; (i = 1,2,3).
\end{equation}

The scalar potential for this sector:
\begin{eqnarray}
V_{triplet}&=& \sum_{i =1}^3 {m^2_{\Delta^L_i}} ~Q_{i}^{tt}
+ m_{\hat\Delta^L}^2 ~O_{1}^{tt}  
+{1 \over 2} \sum_{i =1}^3 \lambda^t_{1_i}
\left[Q_{i}^{tt}\right]^2
+{1 \over 2} \sum_{k < j,\, k =1}^2 \sum_{j =2}^3 
\lambda^t_{2jk} Q_{j}^{tt}Q_{k}^{tt}
\nonumber\\
&+&{1 \over 2} \lambda^t_3 \left \{[O_{1}^{tt}]^2  +
\{O_{2}^{tt}\}^\dagger O_{2}^{tt}  +
O_{1}({T_s^{tt}}, T_s^{tt\dagger}) \right \}
+ {1 \over 2}\sum_{i =1}^3 \lambda^t_{4_i}\left[Q_{i}^{tt} O_{1}^{tt} \right]
\nonumber\\
&+&\lambda^t_5 \left[ \mathscr{O}_{1}^{tt} \Delta_1^L + ~{\rm h.c.}\right]  
+ \lambda^t_6 \left[ \mathscr{O}_{3}^{tt} \Delta_2^L + ~{\rm h.c.}\right]  
+ \lambda^t_7 \left[ \mathscr{O}_{2}^{tt} \Delta_3^L + ~{\rm h.c.}\right]  
\nonumber\\
&+& \sum_{i =1}^3 \lambda^t_{8_i} 
\left[\widetilde{O}_{i}^{tt} \Delta_i^L \Delta_i^L +
~{\rm h.c.} \right]
+   \left[ \lambda^t_{9_1} \widetilde{O}_{1}^{tt}
\Delta_2^L \Delta_3^L + ~{\rm h.c.} + ~{\rm cyclic} \right] \;\;.
\label{Vt}
\end{eqnarray}

\subsection{Inter-sector terms:}

So far, we have listed the terms in the potential that involve
scalar fields which belong to any one of three sectors: singlets,
doublets, or triplets of $SU(2)_L$. Besides these, there will
also be terms in the scalar potential which involve fields from
multiple sectors. Below we list the terms  which
arise from couplings of the singlet sector with either the
doublet or the triplet sector.  The other inter-sector terms --
doublet-triplet type -- are
dropped.  Since the vacuum expectation values of the singlet
fields are by far the largest this is not an unreasonable
approximation.

\subsubsection{Inter-sector Singlet-Doublet terms:}

It is useful to define,
\begin{equation}
\widetilde{T}_{s}^{ss} \equiv T_{s}(\hat{\Delta}^{R},
\hat{\Delta}^{R\dagger}), \;\; {\rm and} \;\; \widetilde{T}_{s}^{dd}
\equiv T_{s}(\Phi, \Phi^{\dagger}),
\end{equation}
and
\begin{equation}
O_{1S}^{sd} \equiv O_{1}(\widetilde{T}_s^{dd},\widetilde{T}_s^{ss});
~\mathscr{O}_{3}^{sd} \equiv O_{3}(\hat{\Delta}^{R},
\widetilde{T}_s^{dd}) \;\;.
\end{equation}
For simplicity, we do not keep the combinations
$\widetilde{T}_{a}^{ss} \equiv T_{a}(\hat{\Delta}^{R},
\hat{\Delta}^{R\dagger})$  and $\widetilde{T}_{a}^{dd}
\equiv T_{a}(\Phi, \Phi^{\dagger})$.

In terms of the above:
\begin{eqnarray}
V_{sd}&=& {1 \over 2} \lambda^{sd}_1\left[Q_3^{ss} Q_{\eta}^{dd} \right]
+ {1 \over 2} \lambda^{sd}_2 \left[Q_3^{ss} O_{1}^{dd}\right] 
+ {1 \over 2} \lambda^{sd}_3\left[Q_{\eta}^{dd} O_{1}^{ss}\right]
+\lambda^{sd}_4\left[\{\mathscr{O}_{3}^{sd}\}^\dagger\Delta_3^R
+ h.c. \right]
\nonumber \\ 
&+& {1 \over 2} \lambda^{sd}_5\left[O_{1}^{dd}O_{1}^{ss} +
\{O_{2}^{ss}\}^\dagger O_{2}^{dd} + \{O_{2}^{dd}\}^\dagger
O_{2}^{ss} +O_{1S}^{sd}\right].
\label{Vsd}
\end{eqnarray}
Here, in the last term, we have made the simplifying assumption
that there is a common coupling $\lambda_5^{sd}$ for the terms in the
potential which arise from various combinations of $(\Phi^\dagger \Phi)
(\hat \Delta^{R\dagger} \hat \Delta^R)$, each of the four fields being $A4$ triplets.

\subsubsection{Inter-sector Singlet-Triplet terms:}

For this case the following combinations arise:
\begin{eqnarray}
O_{i}^{ts} &\equiv&
O_{i}(\hat{\Delta}^{R\dagger},\hat\Delta^L)\;\;(i = 1,2,3); 
O_{1S}^{ts} \equiv O_{1}(\widetilde{T}_s^{tt},\widetilde{T}_s^{ss});
\nonumber\\
\mathscr{O}_{i}^{ts}  &\equiv&
O_{i}(\widetilde{T}_s^{ss},\hat\Delta^L) \;\;(i = 1,2,3); 
\mathscr{\widetilde{O}}_{3}^{ts}  \equiv
O_{3}(\widetilde{T}_s^{tt},\hat\Delta^R) \;\;.
\end{eqnarray}
In line with the convention introduced earlier:
$\widetilde{O}_{i}^{ts}  \equiv
O_{i}(\hat{\Delta}^{R\dagger},\hat\Delta^{L\dagger})\;\;(i = 1, 2, 3)$.

The intersector potential for this case is given by:
\begin{eqnarray}
V_{ts}&=& {1 \over 2} \sum_{i =1}^3 \lambda^{ts}_{1_i}
\left[Q_3^{ss}  Q_{i}^{tt} \right]
+ {1 \over 2} \lambda^{ts}_2 \left[Q_3^{ss} O_{1}^{tt}\right]
+ {1 \over 2} \sum_{i =1}^3 \lambda^{ts}_{3_i}
\left[Q_{i}^{tt} O_{1}^{ss} \right]
\nonumber \\ 
&+&{1 \over 2} \lambda^{ts}_4
\left[O_{1}^{tt}O_{1}^{ss}+\{O_{2}^{ss}\}^\dagger
O_{2}^{tt}
+\{O_{2}^{tt}\}^\dagger O_{2}^{ss}+O_{1S}^{ts}  \right]
\nonumber \\ 
&+&
\sum_{i =1}^3 \lambda^{ts}_{5_i}\left[
\mathscr{O}_{i}^{ts}{\Delta^L_i}^\dagger + h.c. \right]
+ \lambda^{ts}_6\left[\mathscr{\widetilde{O}}_{3}^{ts}
{\Delta_3^R}^\dagger + h.c. \right]
\nonumber \\ 
&+& \lambda^{ts}_7\left[O_{1}^{ts} {\Delta^L_3}^\dagger
\Delta_3^R  + h.c.\right]
+ \lambda^{ts}_8\left[O_{2}^{ts}
{\Delta^L_1}^\dagger   \Delta_3^R + h.c. \right]
+\lambda^{ts}_9\left[O_{3}^{ts} {{\Delta^L_2}^\dagger} \Delta_3^R + h.c.\right]
\nonumber \\ 
&+& \lambda^{ts}_{10}\left[\widetilde{O}_{3}^{ts}
{\Delta_3^R \Delta^L_1} + h.c. \right]
+ \lambda^{ts}_{11}\left[\widetilde{O}_{2}^{ts}
{\Delta_3^R \Delta^L_3} + h.c. \right]
+ \lambda^{ts}_{12}\left[\widetilde{O}_{1}^{ts}
\Delta_3^R \Delta^L_2 + h.c. \right]\;\;.
\label{Vts}
\end{eqnarray}

\subsection{The minimization conditions:}

After having presented the scalar potential we now seek to
find the conditions under which the {\em vev} we have used in the
model -- see eqs. (\ref{vev1}) and (\ref{vev2}) and Table
\ref{tab1s} -- constitute a local minimum. For ready reference
the {\em vev} are:
\begin{equation}
\langle \Phi^0 \rangle = \frac{v}{\sqrt{3}} \pmatrix{1 \cr 1 \cr 1} \;,\;
\langle \hat{\Delta}^{L0} \rangle = v_L\pmatrix{1 \cr 0 \cr 0} \;,\; 
\langle \hat{\Delta}^{R0} \rangle = v_R\pmatrix{1 \cr \omega^2
\cr \omega} \;,\;
\label{vev1a}
\end{equation}
\begin{equation}
\langle \eta^0 \rangle = u  \;,\; 
\langle \Delta_1^{L0} \rangle =  
\langle \Delta_2^{L0} \rangle =  
\langle \Delta_3^{L0} \rangle = {u}_L \;, 
\langle \Delta_3^{R0} \rangle = {u}_R \;.\; 
\label{vev2a}
\end{equation}
where the $SU(2)_L$ nature of the fields is suppressed.

It can be seen from  eq.
(\ref{vev1a}) that the $A4$ triplet fields --
$\hat{\Delta}^{L,R}$ and $\Phi$ -- acquire {\em vev}  which have
been shown to be global minima in \cite{gmin1}. While this is
certainly encouraging,  that result is for one $A4$ triplet in
isolation. Here there are many other fields and so it is not
straight-forward to directly extend the results of \cite{gmin1}. 

In the following we list, sector by sector, the conditions under
which the {\em vev} in eqs. (\ref{vev1a}) and (\ref{vev2a})
correspond to a minimum.

\subsubsection{$SU(2)_L$ singlet sector:}

The {\em vev} of the singlet fields $\hat\Delta^R_i$ and
$\Delta^R_3$ are much larger than those of the $SU(2)_L$ doublet
and triplet scalars. So, the contributions to the minimization
equations from the inter-sector terms can be neglected.

Using the singlet sector potential in eq. (\ref{Vs}) and the {\em
vev} in eqs.  (\ref{vev1a}) and (\ref{vev2a}) we get (bearing in mind
$v_R$ is real):
\begin{equation}
\frac{\partial V_{singlet}|_{min}}{\partial u_R^*}=0
\Rightarrow u_R\left[m_{\Delta_3^R}^2+\lambda^s_1
u_R^*u_R+{3\over2}\lambda^s_3 v_R^2\right]+ 3
v_R^2\left[\lambda^s_4 v_R + 2 \lambda^s_5u_R^* \right]=0 \;\;,
\end{equation}
and
\begin{eqnarray}
\frac{\partial V_{singlet}|_{min}}{\partial v_{Ri}^* }&=& 0
\nonumber \\
& \Rightarrow & v_{R} \left[m_{\hat\Delta^R}^2+ 4\lambda^s_2 v_R^2 
+{\lambda^s_3 \over2}u_R^*u_R+\lambda^s_4
v_R(2u_R+u_R^*)+ 2 \lambda^s_5 u_R^2 \right] =0 \;\;.
\end{eqnarray}

\subsubsection{$SU(2)_L$ doublet sector:}
In this sector we have to include the contributions from both the
doublet sector itself -- eq. (\ref{Vd}) -- as well as the
inter-sector terms in eq. (\ref{Vsd}).  We define
$V_{\mathscr{D}}=V_{doublet}+V_{sd}$.

In order that the potential minimum corresponds to the {\em vev}
in eqs. (\ref{vev1a}) and (\ref{vev2a}) we must have:

\begin{equation}
\frac{\partial V_{\mathscr{D}}|_{min}}{\partial u^*}=0
\Rightarrow u \left[2m_{\eta}^2+2\lambda^d_1 u^*u +\lambda^d_3 v^*v
+\lambda^{sd}_1 u_R^*u_R +3\lambda^{sd}_3 v_R^2 \right]
=0.
\label{DMIN2}
\end{equation}
and
\begin{eqnarray}
\frac{\partial V_{\mathscr{D}}|_{min}}{\partial v_i^* }&=&0
\nonumber\\
&\Rightarrow& \frac{v}{\sqrt{3}} \left[m_{\Phi}^2
+4 \lambda^d_2 \left({{v^*v}\over 3}\right)
 +\lambda^d_3 u^*u + {1\over2}\lambda^{sd}_2 u_R^*u_R \right.
\nonumber\\
&+& \left. \lambda^{sd}_4(u_R^*+u_R)
v_R+\frac{5}{4}\lambda^{sd}_5 v_R^2 \right] =0.
\label{DMIN3}
\end{eqnarray}

Notice that one has to resort to some degree of fine-tuning
to satisfy eqs. (\ref{DMIN2}) and (\ref{DMIN3}) which involve
both $SU(2)_L$ doublet and singlet {\em vev} of quite different magnitudes.

\subsubsection{$SU(2)_L$ triplet sector:}

Using eqs. (\ref{Vs}) and (\ref{Vts}) we define
$V_{\mathscr{T}}=V_{triplet}+V_{ts}$.

In this sector there are a plethora of couplings. To ease the
presentation we choose
\begin{eqnarray}
m_{\Delta^L_1}&=&m_{\Delta^L_2}=m_{\Delta^L_3}=m_{\Delta^L} \;\;; \;\;
\lambda^t_{1_1}=\lambda^t_{1_2}=\lambda^t_{1_3}=\lambda^t_{a}
\;\;; \;\;
\lambda^t_{4_1}=\lambda^t_{4_2}=\lambda^t_{4_3}=\lambda^t_{b}
\nonumber\\
\lambda^t_{221}&=&\lambda^t_{232}=\lambda^t_{231}=\lambda^t_{c} \;\;; \;\;
\lambda^t_{8_1}= \lambda^t_{8_2} = \lambda^t_{8_3} =\lambda^t_{d}
\;\;; \;\;
\lambda^t_{9_1}= \lambda^t_{9_2} = \lambda^t_{9_3} =\lambda^t_{e}\nonumber\\
\lambda^{ts}_{1_1}&=&\lambda^{ts}_{1_2}=\lambda^{ts}_{1_3}=\lambda^{ts}_{a}
\;\;; \;\;
\lambda^{ts}_{3_1}=\lambda^{ts}_{3_2}=\lambda^{ts}_{3_3}=\lambda^{ts}_{b}\;\;;
\;\; 
\lambda^{ts}_{5_1}=\lambda^{ts}_{5_2}=\lambda^{ts}_{5_3}=
\lambda^{ts}_{c}\nonumber\\
\lambda^{ts}_{10}&=&\lambda^{ts}_{11}=\lambda^{ts}_{12}=\lambda^{ts}_{d}
\;\;; \;\;
\lambda^{ts}_7 =\lambda^{ts}_8=\lambda^{ts}_9=\lambda^{ts}_{f}.
\label{TMIN1}
\end{eqnarray}
For the minimization of $V_{\mathscr{T}}$ so as to arrive at the
{\em vev} in eqs.  (\ref{vev1a}) and (\ref{vev2a}) one must
satisfy:
\begin{eqnarray}
\frac{\partial V_{\mathscr{T}}|_{min}}{\partial u_L^*}&=&0\nonumber\\
&\Rightarrow& u_L\left[m^2_{\Delta^L}+
(\lambda^{t}_{a}+\lambda^{t}_{c}) u_L^*u_L  +{1\over2}\lambda^{t}_{b} v_L^*
v_L +
{1\over2}\lambda^{ts}_{a} u_R^*u_R + {3\over2}\lambda^{ts}_{b}
v_R^2\right]\nonumber\\ &+& 2
v_L^2u_L^*(\lambda^t_{d}+\lambda^t_{e}) + 
v_L v_R\left[-{1\over2} \lambda^{ts}_{c}
v_R  + \lambda^{ts}_{d}u_R^*+\lambda^{ts}_{f} u_R \right] =0.
\label{DMIN22}
\end{eqnarray}

Again:
\begin{eqnarray}
\frac{\partial V_{\mathscr{T}}|_{min}}{\partial v_{L1}^*}&=&0\nonumber\\
&\Rightarrow&  v_L\left[m_{\hat\Delta^L}^2 +
{3\over2}\lambda^{t}_{b} u_L^*u_L
+ 2\lambda^{t}_{3} v_L^* v_L +
{1\over2}\lambda^{ts}_{2} u_R^*u_R +{3\over2}\lambda^{ts}_{4}
v_R^2 \right]\nonumber\\ 
&+&u_L\left[6 u_L v_L^* 
(\lambda^{t}_{d}+\lambda^{t}_{e}) -{3\over2}\lambda^{ts}_{c} v_R^2  +
3\lambda^{ts}_{f} u_R^* v_R +3\lambda^{ts}_{d} u_R v_R  \right]
=0.
\label{DMIN32}
\end{eqnarray}

Also we have 
\begin{eqnarray}
\frac{\partial V_{\mathscr{T}}|_{min}}{\partial v_{L2}^*}&=&
\frac{\partial V_{\mathscr{T}}|_{min}}{\partial v_{L3}^*}=0\nonumber\\
&\Rightarrow& v_L v_R \left[ -{1\over4} \lambda^{ts}_4 v_R +
\lambda^{ts}_6 (u_R^*+u_R)\right] =0.
\label{DMIN4}
\end{eqnarray}
Here again fine-tuning is required to ensure that 
eqs. (\ref{DMIN22}) - (\ref{DMIN4}) are satisfied.



\end{document}